\documentstyle[12pt]{article}

\begin{document}
 \thispagestyle{empty}
 \begin{flushright}
 {UTAS-PHYS--97-12}\\[3mm]
 {hep-th/9709216}\\[5mm]
 {September 1997}\\
\end{flushright}
 \vspace*{3cm}
 \begin{center}
 {\bf \Large
 A geometrical angle on Feynman integrals}
 \end{center}
 \vspace{1cm}
 \begin{center}
 A.~I.~Davydychev$^{a,b,}$\footnote{davyd@theory.npi.msu.su}
 \ \ and \ \
 R.~Delbourgo$^{a,}$\footnote{Bob.Delbourgo@utas.edu.au}\\
 \vspace{1cm}
$^{a}${\em 
 Physics \ Department, \ \ \ University \ of \ Tasmania, \\
 GPO Box 252-21, Hobart, Tasmania, 7001 Australia}
\\
$^{b}${\em
 Institute for Nuclear Physics, Moscow State University, \\
 119899, Moscow, Russia}
\end{center}
 \hspace{3in}
 \begin{abstract}
A direct link between a one-loop $N$-point Feynman diagram
and a geometrical representation based on the $N$-dimensional 
simplex is established by relating the Feynman parametric 
representations to the integrals over contents of $(N-1)$-dimensional 
simplices in non-Euclidean geometry of constant curvature.
In particular, the four-point function in four dimensions 
is proportional to the volume of a three-dimensional 
spherical (or hyperbolic) tetrahedron which can be calculated
by splitting into birectangular ones.
It is also shown that the  known formula of reduction of the
$N$-point function in $(N-1)$ dimensions corresponds to splitting 
the related $N$-dimensional simplex into $N$ rectangular ones.
 \end{abstract}

\vspace{7mm}

\begin{center}
{\sf Published in J.~Math.~Phys. 39 (1998) 4299--4334} 
\end{center}

\newpage

\section{Introduction}
\setcounter{equation}{0}

The development of techniques for efficient calculation of one-loop
$N$-point Feynman diagrams is very important for studying
the leading and next-to-leading corrections to elementary
particle processes within and beyond the Standard Model.
Since any one-loop diagram can be reduced to a combination
of scalar integrals \cite{BF+PV},
in what follows we shall mainly deal with scalar Feynman integrals.

It is well known (see e.g. in \cite{'tHV-79}) that results
for the three- and four-point functions in four dimensions
can be expressed in terms of dilogarithms (or related functions).
The first explicit calculation of the general one-loop
four-point function was given in \cite{Wu} and later on, 
more compact results were presented in refs.~\cite{'tHV-79,DNS}
(see also in \cite{vOV,Denner}).
Moreover, it is known that (in four dimensions) the five-point
function, the ``pentagon'', can be reduced to a linear combination
of four-point functions \cite{Halpern,Petersson,Melrose}
(see also in \cite{KT,Nickel,vNV,BDK}).
Using linear dependence of the external momenta, a similar 
reduction procedure can be applied to the $N$-point functions
with $N\geq 6$ (see in \cite{Brown,Petersson}).

As a rule, explicit results for diagrams with several external legs 
possess a rather complicated analytical structure. 
For example, separate terms in a sum of dilogarithms
may have singularities (cuts) which cancel in the whole sum.
As the result, there are certain difficulties in describing
analytic continuation to all regions (in external momenta and masses)
of interest.

Another approach to the evaluation of one-loop integrals \cite{series}
makes it possible to represent them in terms of multiple hypergeometric
functions. In ref.~\cite{jmp}, such results for the integrals with an
arbitrary number of external legs were presented. As a rule, the 
corresponding hypergeometric series have a rather restricted region of convergence. In general, the problem of analytic continuation of 
occurring functions to all other regions of interest is very complicated.

The analytical structure of the results can be better understood
if one employs a geometrical interpretation of kinematic invariants
and other quantities. For example, the singularities of 
the general three-point function can be described pictorially through a
tetrahedron constructed out of the external and internal momenta.
This method can be used to derive Landau equations defining the 
positions of possible singularities \cite{Landau} (see
also in \cite{KW,Mandelstam,ELOP}) and a 
similar approach can be applied to the four-point function
\cite{KSW2,Wu} too.  

Another well-known example of using geometrical ideas
is the massless three-point function with arbitrary (off-shell)
external momenta. Let us denote them as $k_{12}, \; k_{23}$ and 
$k_{31}$, so that $k_{12}+k_{23}+k_{31}=0$, and assume that
all these momenta are time-like ($k_{jl}^2>0$). 
Using the standard notation for the ``triangle'' K\"{a}llen function,
\begin{equation}
\label{Kallen}
\lambda(x,y,z)= x^2 + y^2 + z^2 - 2xy - 2yz - 2zx ,
\end{equation}
and assuming that we are in the region 
$\lambda\left(k_{12}^2,k_{23}^2,k_{31}^2\right)\leq~0$, the result
for the corresponding one-loop scalar Feynman 
integral\footnote{Here we use
the same normalization as in (\ref{sun}). The results for the 
massless three-point functions can be found e.g. 
in \cite{BC2,'tHV-79,jpa}. 
They are closely connected with the results for two-loop massive 
vacuum diagrams \cite{2loop_vacuum,Scharf,magic}.}
(in four dimensions) can be neatly expressed as\footnote{The 
functions related to the dilogarithm (including the Clausen function
$\mbox{Cl}_2(\theta)$) are defined in Appendix~A (see also in
\cite{Lewin}).}
\begin{equation}
\label{massless_triangle}
\frac{2\mbox{i}\pi^2}
     {\sqrt{-\lambda\left(k_{12}^2,k_{23}^2,k_{31}^2\right)}} \;
\left\{ \mbox{Cl}_2\left(2\theta_{12}\right) 
       + \mbox{Cl}_2\left(2\theta_{23}\right)
       + \mbox{Cl}_2\left(2\theta_{31}\right) \right\}
\end{equation}
where the angles $\theta_{12}, \; \theta_{23}$ and $\theta_{31}$
are nothing but those of the triangle with sides 
$\sqrt{k_{12}^2}$, $\sqrt{k_{23}^2}$
and $\sqrt{k_{31}^2}$, as shown in Fig.~1. Moreover, the denominator
$\sqrt{-\lambda}$ is nothing but four times the area of this
triangle. 

In this paper, we discuss the geometrical interpretation of the
kinematic variables related to the one-loop $N$-point functions.
The generalization of the tetrahedron representation (which was
used in the three-point case) is an $N$-dimensional simplex.
Furthermore, we show that there is a direct transition from the
Feynman parametric representation to the geometrical description
connected with an $N$-dimensional simplex.
We thereby arrive at a ``geometrical'' way of evaluating
Feynman integrals. 

The paper is organized as follows. In Section~2 we describe the
connection of the $N$-point variables and the corresponding
$N$-dimensional simplex.  
In Section~3 we show how one can ``geometrize'' the ordinary Feynman
parametric representation. 
In Sections~4,5,6 we consider application of these geometrical
ideas to the two-, three- and four-point functions, respectively.
In Section~7 we discuss how to use geometrical ideas for the
reduction of $N$-point integrals in $(N-1)$ dimensions,
and also for the calculation of the integrals with higher powers
of denominators.
Finally, in the last Section~8, we summarize the main ideas and
results.

\section{Basic simplices in $N$ dimensions}  
\setcounter{equation}{0}

\subsection{$N$-point function and the related simplex}

Consider a one-loop scalar $N$-point ``sun-type'' diagram presented
in Fig.~2. 
The corresponding Feynman integral is
\begin{equation}
\label{sun}
J^{(N)}(n; \nu_1, \ldots, \nu_N )\equiv
\int \frac{\mbox{d}^n q}
{\prod_{i=1}^N \left[\left(p_i+q\right)^2 - m_i^2 \right]^{\nu_i}} ,
\end{equation}
where $n$ is the space-time dimension and $\nu_i$ are the
powers of the propagators\footnote{Here and below, the usual
causal prescription for the propagators is understood,
i.e. $1/\left[q^2-m^2\right]^{\nu}\leftrightarrow 
1/\left[q^2-m^2+\mbox{i}0\right]^{\nu}$.}. In general, it depends on ${\textstyle{1\over2}} N(N-1)$ momenta invariants
$k_{jl}^2$ ($j<l$), where
\begin{equation}
\label{k_p}
k_{jl} \equiv p_j-p_l ,
\end{equation}
and $N$ masses $m_i$ corresponding to the internal propagators.
The momenta $p_i$ are auxiliary in the sense that they all can be
shifted by a constant vector without changing the momenta
$k_{jl}$ (\ref{k_p}). We shall discuss this freedom below
(see eqs.~(\ref{sun2})--(\ref{sun3})).

It is well known that, using the condition $\sum\alpha_i=1$, 
the standard quadratic form occurring 
in the denominator of the integrand of the Feynman
parametric representation for these integrals
(cf. eqs.~(\ref{Fp1})--(\ref{Fp2}) below) can be rewritten 
in a homogeneous form,
\begin{equation}
\label{homo}
\left[
   \begin{array}{c} 
   {} \\[-3mm] {\sum\sum} \\[-3mm] {}_{j<l} 
   \end{array} 
   \alpha_j \alpha_l k_{jl}^2 - \sum \alpha_i m_i^2 
\right]
\Rightarrow  - \left[ \sum \alpha_i^2 m_i^2 \!+\!
2\!\begin{array}{c} {} \\[-3mm] {\sum\sum} \\[-3mm] {}_{j<l} \end{array}
\alpha_j \alpha_l m_j m_l c_{jl} \right] ,
\end{equation}
where\footnote{In some papers, the notation $y_{jl}$ is used for
$\pm c_{jl}$.}
\begin{equation}
\label{def_c}
c_{jl} \equiv \frac{m_j^2 + m_l^2 - k_{jl}^2}{2 m_j m_l} .
\end{equation}
In the region between the corresponding two-particle pseudo-threshold,
$k_{jl}^2=(m_j-m_l)^2$, and the threshold, $k_{jl}^2=(m_j+m_l)^2$,
we have $|c_{jl}|<1$, and therefore in this region they can be 
understood as cosines of some angles $\tau_{jl}$,
\begin{equation}
\label{def_costau}
c_{jl} = \cos\tau_{jl} 
= \left\{ \begin{array}{c} \;\; 1, \;\;\; k_{jl}^2=(m_j-m_l)^2 \\
                               -1, \;\;\; k_{jl}^2=(m_j+m_l)^2 
          \end{array} \right. \; .
\end{equation}
The corresponding angles $\tau_{jl}$ are
\begin{equation}
\label{def_tau}
\tau_{jl}= \arccos(c_{jl}) 
=\arccos\left(\frac{m_j^2+m_l^2-k_{jl}^2}{2m_j m_l}\right)
= \left\{ \begin{array}{c}   0, \;\;\; k_{jl}^2=(m_j-m_l)^2 \\
                           \pi, \;\;\; k_{jl}^2=(m_j+m_l)^2
          \end{array} \right. \; .
\end{equation}

The expressions in other regions should be understood in the sense
of analytic continuation, using (when necessary) the causal prescription 
for the propagators. For example, when $k_{jl}^2<(m_j-m_l)^2$
($c_{jl}>1$), i.e. below the pseudothreshold, we should interpret
\begin{equation}
\tau_{jl}=-\mbox{i}\; \mbox{Arch}(c_{jl})
= -\frac{\mbox{i}}{2} 
\ln\left( 
\frac{m_j^2+m_l^2-k_{jl}^2+\sqrt{\lambda(m_j^2,m_l^2,k_{jl}^2)}}
     {m_j^2+m_l^2-k_{jl}^2-\sqrt{\lambda(m_j^2,m_l^2,k_{jl}^2)}}
\right) ,
\end{equation}
where $\lambda(x,y,z)$ is the K\"{a}llen function
defined by eq.~(\ref{Kallen}). 
When $k_{jl}^2>(m_j+m_l)^2$ ($c_{jl}<-1$), i.e. above the threshold, 
we get
\begin{equation}
\tau_{jl}=\pi+\mbox{i}\; \mbox{Arch}(-c_{jl})
= \pi + \frac{\mbox{i}}{2}
\ln\left(
\frac{k_{jl}^2-m_j^2-m_l^2+\sqrt{\lambda(m_j^2,m_l^2,k_{jl}^2)}}
     {k_{jl}^2-m_j^2-m_l^2-\sqrt{\lambda(m_j^2,m_l^2,k_{jl}^2)}}
\right) .
\end{equation}
Note that $\lambda(m_j^2,m_l^2,k_{jl}^2)$ is positive when
$k_{jl}^2<(m_j-m_l)^2$ or $k_{jl}^2>(m_j+m_l)^2$.

When the angles $\tau_{jl}$ (defined by eq.~(\ref{def_tau})) are real,
the quadratic form on the r.h.s. in eq.~(\ref{homo}) has a rather
simple geometrical interpretation. Let us consider a set of
Euclidean ``mass'' vectors whose lengths are $m_i$. Let them be directed
so that the angle between the $j$-th and the $l$-th vectors
is $\tau_{jl}$. If we denote
the corresponding unit vectors as $a_i$ (so that the ``mass'' vectors
are $m_i a_i$), we get
\begin{equation}
(a_j \cdot a_l) = \cos\tau_{jl} = c_{jl}. 
\end{equation}
This is also valid for the case $j=l$, since $\tau_{jj}=0$ and
$c_{jj}=1$, cf. eqs.~(\ref{def_costau})--(\ref{def_tau}).
Now, we can represent eq.~(\ref{homo}) as
\begin{equation}
 \sum \alpha_i^2 m_i^2 \!+\!
2\!\begin{array}{c} {} \\[-3mm] {\sum\sum} \\[-3mm] {}_{j<l} \end{array}
\alpha_j \alpha_l m_j m_l c_{jl}
= \left( \sum \alpha_i m_i a_i \right)^2 .
\end{equation}

In $N$ dimensions, if we put all ``mass'' vectors together as emanating
from a common origin, they, together with the sides connecting
their ends, will define a {\em simplex} (or hyper-tetrahedron) which is 
the {\em basic} one for a given Feynman diagram. 
An example of such a simplex for $N=4$ is illustrated by 
Fig.~3.\footnote{Since the four-dimensional Euclidean space is
understood, one should consider Fig.~3 just as an illustration
(rather than precise picture).}
It is easy to see that
the length of the side connecting the ends of the $j$-th and the $l$-th
mass vectors is $\left(m_j^2+m_l^2-2m_j m_l c_{jl}\right)^{1/2}
= \left(k_{jl}^2\right)^{1/2}$, so we shall call it a ``momentum'' side. 
In total, the {\em basic} $N$-dimensional simplex 
has ${\textstyle{1\over2}}N(N+1)$ sides, 
among them $N$ mass sides (corresponding to the masses 
$m_1, \ldots, m_N$) and ${\textstyle{1\over2}}N(N-1)$ 
momentum sides (corresponding to the momenta $k_{jl}, j<l$), 
which meet at $(N+1)$ vertices. 
Each vertex is a ``meeting point'' for $N$ sides. 
There is one vertex where all mass sides meet, which we shall call 
the {\em mass meeting point}, $M$. 
All other vertices are meeting points for
$(N-1)$ momentum sides and one mass side. 
Furthermore, the number of  $(N-1)$-dimensional hyperfaces is $(N+1)$.
$N$ of them (the {\em reduced} hyperfaces) can be obtained by excluding 
one mass side in turn, together with the corresponding vertex and 
the momentum sides meeting at that vertex. The reduced hyperface 
without the $j$-th mass side 
is nothing but the $(N-1)$-dimensional simplex corresponding 
to an $(N-1)$-point function
obtained from the $N$-point function by shrinking the $j$-th 
propagator; this is equivalent to the power of propagator vanishing, 
$\nu_j=0$. The last hyperface, namely the {\em momentum} hyperface 
involves all momentum sides but does not include any mass sides. 
It is associated with the massless $N$-point function possessing the 
same external momenta as the basic one.

In general, in the $N$-point function only $(N-1)$ external momenta
are independent, because they are related via the conservation law.
We can choose, for example, the vectors
$k_{1N}, k_{2N}, \ldots, k_{N-1,N}$. 
Considering all scalar products of these
momenta, including their squares, we get ${\textstyle{1\over2}}N(N-1)$ 
invariants, as stated above. However, when $n$ is integer and
less than $(N-1)$, only $n$ of the $(N-1)$ momenta are independent,
e.g. $k_{1N}, k_{2N}, \ldots, k_{nN}$. Each of the remaining 
$(N-n-1)$ vectors $k_{jN}$ ($j=n+1,\ldots,N-1$) can be fixed
by $n$ scalar products $(k_{jN}\cdot k_{lN})$ ($l=1,\ldots,N-1$).
Hence the total number of independent momentum invariants
for integer $n\leq N-1$ is 
\begin{equation}
{\textstyle{1\over2}}n(n+1) + n(N-n-1)= {\textstyle{1\over2}}n(2N-n-1) ,
\end{equation}   
while for $n\geq N-1$ it remains equal to 
${\textstyle{1\over2}}N(N-1)$; both expressions coincide for $n=N-1$.
The linear dependence between the external momenta reveals itself
in the degeneracy of the results associated with the vanishing
of certain hypervolumes (see below). 

We note that there is a difference between the ``real'' momenta
of the $N$-point diagram, $k_{jl}$, and the vectors
\begin{equation}
K_{jl}\equiv m_j a_j - m_l a_l ,
\end{equation}
corresponding to the momentum sides of the simplex
(the same applies to the momenta $p_i$ and $P_i$ discussed
in subsection~2.4).
The momenta $k_{jl}$ are defined in {\em pseudo-Euclidean}
(Minkowski) space, and they are time-like ($k_{jl}^2>0$)
in the region between the threshold and pseudo-threshold. 
The momentum sides of the simplex, $K_{jl}$, are vectors 
in {\em Euclidean} space, the (Euclidean)
squares of their lengths ($K_{jl}^2$) being equal to $k_{jl}^2$. 
In what follows, we shall consider the $K_{jl}$ as 
Euclidean analogues of the $k_{jl}$ and, vice versa,
the $k_{jl}$ are pseudo-Euclidean analogues of the $K_{jl}$. 
Since the $k_{jl}^2$ ($j<l$) form the complete basis of 
external invariants, all invariant scalars in Minkowski and 
Euclidean spaces are the same.

The matrix with the components (\ref{def_c}), 
\begin{equation}
\label{c-matrix}
\| c \| \equiv
\| c_{jl} \| \equiv
\left(
\begin{array}{c}
{ \; 1\;\;  \;\; c_{12}\; \;\; c_{13}\; \;\; \ldots \;\; c_{1N}   } \\
{ c_{12}\;  \;\; \; 1\;\; \;\; c_{23}\; \;\; \ldots \;\; c_{2N}   } \\
{ c_{13}\;  \;\; c_{23}\; \;\; \; 1\;\; \;\; \ldots \;\; c_{3N}   } \\
{ \ldots\ldots\ldots\ldots\ldots\ldots\ldots } \\
{ c_{1N}    \;\; c_{2N}   \;\; c_{3N}   \;\; \ldots \;\; \; 1\;\; }
\end{array}
\right) \;\; ,
\end{equation}
is associated with many geometrical properties of the 
basic simplex\footnote{The matrix (\ref{c-matrix}) is nothing but
the Gram matrix of the vectors $a_1,\ldots,a_N$.}.
In particular, we shall need its determinant,
\begin{equation}
\label{D(N)}
D^{(N)} \equiv \det\|c_{jl}\| ,
\end{equation}
and the minors
\begin{equation}
\label{minors}
D_{jl}^{(N-1)} \equiv \left\{ \mbox{minor of (\ref{c-matrix})
obtained by eliminating the $j$-th row
and the $l$-th column} \right\} \; .
\end{equation}
In particular, when $j=l$ the minor $D_{jj}^{(N-1)}$ is
a {\em principal} one for the matrix (\ref{c-matrix}).

The formulae for the {\em content} or hypervolume of the
basic $N$-dimensional simplex and its hyperfaces,
in terms of the determinant 
(\ref{D(N)}) and the minors (\ref{minors}), are well known
in the $N$-dimensional Euclidean geometry (see for 
instance \cite{Sommerville,Kendall}).
Thus the content of the $N$-dimensional simplex is given by
\begin{equation}
\label{V^{(N)}}
V^{(N)} = \frac{1}{N!} \left(\prod\limits_{i=1}^N m_i \right)
\sqrt{D^{(N)}} \;\; .
\end{equation}
The content of the $j$-th $(N-1)$-dimensional reduced hyperface
(which does not contain the $j$-th mass side) is
\begin{equation}
\label{V_j}
\overline{V}_j^{(N-1)} = \frac{1}{(N-1)!}
\left(\prod\limits_{i\neq j}^{} m_i \right)
\sqrt{D_{jj}^{(N-1)}} \;\; ,
\end{equation}
whereas the content of the $(N-1)$-dimensional momentum hyperface is
\begin{equation}
\label{V_0}
\overline{V}_0^{(N-1)} = \frac{1}{(N-1)!}
\sqrt{\Lambda^{(N)}} ,
\end{equation}
where $\Lambda^{(N)}$ is symmetric combination of the momenta
defined in the same way as in \cite{Nickel},
\begin{equation}
\label{Lambda(N)}
\Lambda^{(N)} \equiv \det \| (k_{jN} \cdot k_{lN}) \| .
\end{equation}
Note that the content of the basic simplex can also be presented as
\begin{equation}
V^{(N)} = \frac{1}{N} \; \overline{V}_0^{(N-1)} \; m_0 ,
\end{equation}
where $m_0$
is the distance between the mass meeting 
point and the momentum hyperface, i.e. the length of the vector
of the height of the simplex, $H_0$.
Therefore,
\begin{equation}
\label{m_0}
m_0 \equiv \left|H_0\right| 
= N \; \frac{V^{(N)}}{\overline{V}_0^{(N-1)}}
= \left( \prod\limits_{i=1}^N m_i \right) \;
\sqrt{ \frac{D^{(N)}}{\Lambda^{(N)}} } .
\end{equation}

\subsection{Normals, dihedral angles and the dual matrix}

The dihedral angles between the hyperfaces can be defined via the
angles between their normals. Thus, we get 
${\textstyle{1\over2}}N(N-1)$ dihedral angles between pairs of 
the reduced hyperfaces and $N$ angles between the momentum 
hyperface and the reduced ones. 

Again, let $a_i$ be unit vectors directed along the mass sides of the
basic $N$-dimensional simplex. The normals to the $(N-1)$-dimensional
reduced hyperfaces can be defined as
\begin{equation}
\label{n_j}
n_{j \lambda} = \frac{\partial}{\partial a_j^{\lambda}}
\left(
\epsilon_{\nu_1 \ldots \nu_N} \;
  a_{1}^{\nu_1} \ldots a_{N}^{\nu_N} \right) .
\end{equation}
where $\epsilon_{\nu_1 \ldots \nu_N}$ is the completely antisymmetric 
tensor in $N$ dimensions. 

Using the well-known expression for the contraction of the product
of two $\epsilon$ tensors in terms of the determinant with
Kronecker $\delta$ symbols, it is easy to show that
\begin{equation}
\label{(njnl)}
(n_j \cdot n_l) = 
(-1)^{j+l} \; D_{jl}^{(N-1)},
\end{equation}
where $D_{jl}^{(N-1)}$ is defined in (\ref{minors}).
In particular, $n_j^2$
is nothing but the principal minor $D_{jj}^{(N-1)}$
related to the $(N-1)$-point function associated with the
corresponding reduced hyperface. The cosine of the angle 
$\widetilde{\tau}_{jl}$ 
between the normals to the $j$-th and the $l$-th reduced hyperfaces is
\begin{equation}
\label{cos_psi}
\widetilde{c}_{jl} 
\equiv \cos\widetilde{\tau}_{jl}
= \frac{(n_j \cdot n_l)}{\sqrt{n_j^2 n_l^2}}
= \frac{ (-1)^{j+l} \; D_{jl}^{(N-1)}}
       {\sqrt{D_{jj}^{(N-1)} \; D_{ll}^{(N-1)}}} .
\end{equation}
In particular, this means that the matrix with the elements
$\sqrt{D_{jj}^{(N-1)} \; D_{ll}^{(N-1)}}\;\widetilde{c}_{jl}/D^{(N)}$
is the inverse of the matrix (\ref{c-matrix}). 
Namely,
\begin{equation}
\label{inverse}
\| c_{jl} \|^{-1} = \frac{1}{D^{(N)}} \;
\mbox{diag}\left(\sqrt{D_{11}^{(N-1)}}, \ldots, 
                 \sqrt{D_{NN}^{(N-1)}}\right) \;
\| \widetilde{c}_{jl} \| \;
\mbox{diag}\left(\sqrt{D_{11}^{(N-1)}}, \ldots, 
                 \sqrt{D_{NN}^{(N-1)}}\right) ,
\end{equation}
where $D^{(N)}$ is defined in (\ref{D(N)}). We shall call the matrix
$\|\widetilde{c}_{jl}\|$ the  {\em dual}  matrix\footnote{The matrix
$\|\widetilde{c}_{jl}\|$ is nothing but the Gram matrix of unit 
vectors directed along the normals $n_i$ ($i=1,\ldots,N$).} 
(with respect to $\|c_{jl}\|$).
A trivial corollary of (\ref{inverse}) is that
\begin{equation}
\label{D-tilde}
\widetilde{D}^{(N)} \equiv
\det\| \widetilde{c}_{jl} \| 
= \frac{\left(D^{(N)}\right)^{N-1}}
       {\prod_{i=1}^{N} D_{ii}^{(N-1)}} .
\end{equation}
Note that the dihedral angles $\psi_{jl}$ are related
to the angles $\widetilde{\tau}_{jl}$ as
\begin{equation}
\label{psi_bar}
\psi_{jl} = \pi - \widetilde{\tau}_{jl} ,
\hspace{14mm}
\cos\psi_{jl} = -\cos\widetilde{\tau}_{jl} .
\end{equation}

The normal to the momentum hyperface can be defined as
\begin{equation}
\label{n_0}
n_{0 \lambda} = -\epsilon_{\mu_1 \ldots \mu_{N-1} \lambda} \;
\left(K_{1N}\right)^{\mu_1} \ldots
\left(K_{N-1,N}\right)^{\mu_{N-1}} .
\end{equation}
If we represent 
$K_{jN}=(m_j a_j - m_N a_N)$
and use the fact that all terms involving two or more $m_N a_N$
disappear in (\ref{n_0}) (due to the antisymmetry of the $\epsilon$ 
tensor), we get the following representation:
\begin{equation}
\label{another_n_0}
n_{0 \lambda} = -\left( \prod\limits_{i=1}^N m_i \right) \;
\sum\limits_{j=1}^N \frac{n_{j \lambda}}{m_j} .
\end{equation}

On one hand, the original definition (\ref{n_0}) leads to
\begin{equation}
\label{n_0^2}
n_0^2 = \det\| (k_{jN} \cdot k_{lN}) \| \equiv \Lambda^{(N)}
\end{equation}
where $\Lambda^{(N)}$ is 
defined in (\ref{Lambda(N)}).
On the other hand, using (\ref{another_n_0}) we obtain
\begin{equation}
\label{another_n_0^2}
n_0^2 = \left( \prod\limits_{i=1}^N m_i^2 \right) \;
\sum\limits_{l=1}^N \frac{1}{m_l^2} \; F_l^{(N)}
\end{equation}
where (cf. eq.~(4) of \cite{Nickel})
\begin{equation}
\label{F_l}
F_l^{(N)} = 
\sum\limits_{j=0}^N (-1)^{j+l} \; D_{jl}^{(N-1)} \;
\frac{m_l}{m_j}
= \frac{\partial}{\partial m_l^2} \left( m_l^2 D^{(N)} \right) , 
\end{equation}
i.e. $F_l^{(N)}$ is the determinant obtained from $\det\|c_{jl}\|$
by substituting, instead of the $l$-th column, the mass
ratios $m_l/m_j$ (where $j$ is the line number). 
When taking the partial derivative in (\ref{F_l}), it is implied that 
the set of independent variables involves $m_i^2$ and $k_{jl}^2$, rather
than $c_{jl}$. 
In particular, the representation of $F_l^{(N)}$ in terms 
of determinants (\ref{F_l}) (see also in \cite{Nickel}) 
shows that they obey
the following set of linear equations:
\begin{equation}
\label{linear}
\sum_{l=1}^{N} c_{jl} \; F_l^{(N)} \; \frac{1}{m_l}
= D^{(N)} \; \frac{1}{m_j} .
\end{equation}

Another useful representation of $n_0$ is
\begin{equation}
\label{n_0_a}
n_0 = -\frac{\prod m_l}{\sqrt{D^{(N)}}} \;
\sum\limits_{i=1}^N \frac{1}{m_i} F_i^{(N)} a_i , 
\end{equation}
where, as usually, $a_i$ are the unit vectors directed along the mass 
sides of the basic simplex. 
Using eq.~(\ref{linear}) (and remembering that 
$(a_j\cdot a_l)=c_{jl}$), one can easily see that the representation
(\ref{n_0_a}) is equivalent to eqs.~(\ref{n_0}) and
(\ref{another_n_0}). Namely, $n_0$ (defined by (\ref{n_0_a}))
is orthogonal to all $K_{jl}=m_ja_j-m_la_l$ and $n_0^2$ is the
same as in eq.~(\ref{another_n_0^2}).

\subsection{Splitting the basic simplex}

The geometrical meaning of $F_l^{(N)}$ can be understood as follows.
Using eqs.~(\ref{another_n_0}) and (\ref{(njnl)}), we get
\begin{equation}
(n_0 \cdot n_l) =
-\left( \prod\limits_{i\neq l}^{} m_i \right) \;
F_l^{(N)} .
\end{equation}
Therefore, the cosine of the angle $\widetilde{\tau}_{0l}$ 
between the normals to the
momentum hyperface and the $l$-th reduced hyperface is
\begin{equation}
\label{cos_chi_j}
\widetilde{c}_{0l} \equiv \cos\widetilde{\tau}_{0l}
= \frac{(n_0 \cdot n_l)}{\sqrt{n_0^2 n_l^2}}
= -\frac{\left( \prod\limits_{i\neq l}^{} m_i \right) \; F_l^{(N)}}
       {\sqrt{\Lambda^{(N)}\; D_{ll}^{(N-1)}}} ,
\end{equation}
whereas the dihedral angle between these hyperfaces is given by
\begin{equation}
\label{dihedral_0l}
\psi_{0l} = \pi - \widetilde{\tau}_{0l}, \hspace{13mm}
\cos\psi_{0l} = - \cos\widetilde{\tau}_{0l} .
\end{equation}
 
The orthogonal projection of the $i$-th reduced hyperface
onto the momentum hyperface is
\begin{equation}
\label{proj}
\overline{V}_i^{(N-1)} \; \cos\psi_{0i} .
\end{equation}
The sum of all such projections should cover the whole momentum
hyperface,
\begin{equation}
\label{sum_proj_0}
\sum\limits_{i=1}^N \overline{V}_i^{(N-1)} \; 
\cos\psi_{0i} = \overline{V}_0^{(N-1)} .
\end{equation}
Using eqs.~(\ref{cos_chi_j}), (\ref{V_j}) and (\ref{V_0}), 
we see that the condition (\ref{sum_proj_0}) is equivalent to 
\begin{equation}
\label{sum_F_l}
\left( \prod\limits_{j=1}^N m_j^2 \right) \;
\sum\limits_{l=1}^N \frac{F_l^{(N)}}{m_l^2} = \Lambda^{(N)} ,
\end{equation}
which is the case, cf. eqs.~(\ref{n_0^2}), (\ref{another_n_0^2}).
The condition (\ref{sum_F_l}) is equivalent to eq.~(38) of
\cite{Nickel} (see also in \cite{Wu}). Eq.~(\ref{sum_proj_0}) 
illustrates the geometrical meaning of (\ref{sum_F_l}).
Moreover, we can see that, if we use the ``height'' $H_0$ for
splitting the basic $N$-dimensional simplex into $N$ rectangular 
ones whose 
``bases'' correspond to the projections (\ref{proj}), the content of
the $i$-th rectangular simplex $V_i^{(N)}$ is proportional 
to $F_i^{(N)}$, namely
\begin{equation}
\label{V_i^{(N)}}
V_i^{(N)} =
\frac{1}{N}\; \overline{V}_i^{(N-1)} \; m_0 \; \cos\psi_{0i} 
= \frac{V^{(N)}}{\Lambda^{(N)}} \; 
\left( \prod\limits_{l\neq i}^{} m_l^2 \right) \;
F_i^{(N)} .
\end{equation}

In the same manner we can consider a projection onto one of the
reduced hyperfaces. This gives
\begin{equation}
\sum\limits_{l\neq i}^{} \overline{V}_l^{(N-1)} \; 
\cos\psi_{il}
+ \overline{V}_0^{(N-1)} \; \cos\psi_{0i} = \overline{V}_i^{(N-1)} 
\end{equation}
or
\begin{equation}
\label{proj0}
\sum\limits_{l=1}^N \overline{V}_l^{(N-1)} \; 
\widetilde{c}_{il}
+ \overline{V}_0^{(N-1)} \; \widetilde{c}_{0i} =0 .
\end{equation}
In terms of $F_l^{(N)}$, eq.~(\ref{proj0}) corresponds 
to the definition (\ref{F_l}). 

The cosine of the angle $\tau_{0j}$ between the $j$-th mass side
and the height $H_0$ is
\begin{equation}
\label{c0j}
c_{0j} \equiv \cos\tau_{0j} = \frac{m_0}{m_j} .
\end{equation}
Consider the matrix of cosines (similar to (\ref{c-matrix}))
associated with the $i$-th rectangular simplex. Its elements are
\begin{eqnarray}
\label{c-matrix_i}
\left\{
\begin{array}{c}
c_{jl}, \hspace{7mm} \mbox{if} \;\; j\neq i \;\; 
                     \mbox{and} \;\; l\neq i     \\
c_{0l},  \hspace{7mm} \mbox{if} \;\; j=i \;\; \mbox{and} \;\; l\neq i \\
c_{0j},  \hspace{7mm} \mbox{if} \;\; l=i \;\; \mbox{and} \;\; j\neq i \\ 
1,  \hspace{9mm} \mbox{if} \;\; j=l=i \hspace{10mm}                         
\end{array}
\right. 
\end{eqnarray}
Denoting 
\begin{equation}
\label{D_i^{(N)}}
D_i^{(N)} \equiv \left\{ \mbox{determinant of the matrix
(\ref{c-matrix_i})} \right\} ,
\end{equation}
and using eqs.~(\ref{V^{(N)}}) and  (\ref{m_0}), we get
\begin{equation}
\label{V_i^{(N)}-2}
V_i^{(N)} = \left( \prod\limits_{l\neq i}^{} m_l \right)
\sqrt{\frac{D_i^{(N)}}{\Lambda^{(N)}}} .
\end{equation}
Comparing eqs.~(\ref{V_i^{(N)}}) and (\ref{V_i^{(N)}-2}),
we obtain the following result for the determinant (\ref{D_i^{(N)}}): 
\begin{equation}
\label{D_i^{(N)}-2}
D_i^{(N)} = \left( \prod\limits_{l\neq i}^{} m_l \right)
\frac{\left(F_i^{(N)}\right)^2}{\Lambda^{(N)}} .
\end{equation}

\subsection{Choice of the momenta $p_i$}

Let us discuss possibilities for a choice 
of the momenta $p_i$ in eq.~(\ref{sun}) and their Euclidean 
analogues $P_i$, such that $K_{jl}=P_j-P_l$.
Due to eq.~(\ref{k_p}), the $p_i$ vectors emanating from an origin,
together with the vectors $k_{jl}$, form an $N$-dimensional simplex
in pseudo-Euclidean space\footnote{For integer $n<N$, this simplex
is degenerate, since one cannot use more than $n$ dimensions for
external vectors.}.
Analogously, in Euclidean space the vectors $P_i$ 
together with the $K_{jl}$ momentum hyperface (taken as a base), 
form an $N$-dimensional Euclidean simplex similar to the basic one. 
Using translational invariance of $p_i$,
we can shift their origin as we like. 
In particular, we can make one of the $p_i$ vectors zero,
whereupon the origin of $P_i$ coincides with
one of the vertices belonging to the momentum hyperface.
However, this way would break the symmetry of the problem.
There are at least two reasonable, {\em symmetric} ways to
fix the origin of $p_i$ (and $P_i$):
\begin{enumerate}
\item
Choose the mass meeting point as the origin of $P_i$ vectors,
so that $P_i^2=p_i^2=m_i^2$ and the $P_i$ vectors coincide with the
mass sides, $P_i=m_i a_i$. In this case, the basic simplex
becomes even ``more basic'', and the integral (\ref{sun})
can be expressed as
\begin{equation}
\label{sun2}
\left.
J^{(N)}(n; \nu_1, \ldots, \nu_N ) =
\int \frac{\mbox{d}^n q}
{\prod_{i=1}^N \left[q^2 + 2 (p_i\cdot q) 
\right]^{\nu_i}}\right|_{(p_j\cdot p_l)=m_j m_l c_{jl}} .
\end{equation}
Such a choice of $p_i$ simplifies study of singularities\footnote{In
this case, the main Landau singularity is associated with $q\to 0$.}
via Landau equations \cite{Landau,ELOP}.
Note that, according to eq.~(\ref{n_0_a}), in this case the vector 
of the height $H_0$ can be represented as
\begin{equation}
H_0=\frac{\prod m_l^2}{\Lambda^{(N)}} 
\sum\limits_{i=1}^N \frac{1}{m_i^2} F_i^{(N)} P_i .
\end{equation} 
However, this choice of $p_i$ cannot be used for integer $n<N$.

\item
Pick, as the origin of $P_i$ vectors, the point of intercept
of the height $H_0$ and the momentum hyperface.
This is nothing but a projection of the previous option
onto the momentum hyperface. In this case (cf. eq.~(\ref{c0j})),
\begin{eqnarray}
p_i^2 = P_i^2 = m_i^2 - m_0^2 = m_i^2 (1-c_{0i}^2), 
\hspace{30mm} \\
(p_j\cdot p_l) = (P_j\cdot P_l) = m_j m_l c_{jl} - m_0^2
= m_j m_l \left( c_{jl} - c_{0j} c_{0l} \right) ,
\end{eqnarray}
and the denominators in the integral (\ref{sun}) become
\begin{equation}
\label{sun3}
\left[ (p_i+q)^2 - m_i^2 \right]
\Rightarrow
\left[ q^2 + 2 (p_i\cdot q) - m_0^2 \right] .
\end{equation}
The projection of eq.~(\ref{n_0_a}) onto the momentum hyperface
shows that there exists the following relation which is valid
for {\em this} choice of $P_i$: 
\begin{equation}
\sum\limits_{i=1}^N \frac{1}{m_i^2} F_i^{(N)} P_i = 0 .   
\end{equation}

\end{enumerate}

\section{From Feynman parameters to non-Euclidean geometry}
\setcounter{equation}{0}

\subsection{Moving from linear to quadratic hypersurface}

The standard Feynman parametric representation of a one-loop 
$N$-point integral in $n$ dimensions (\ref{sun}),
corresponding to the diagram presented in Fig.~2, reads
\begin{equation}
\label{Fp1}
J^{(N)}\left(n; \nu_1, \ldots , \nu_N \right)
= \mbox{i}^{1-n} \pi^{n/2} 
\frac{\Gamma\left(\sum\nu_i \!-\! {\textstyle{n\over2}} \right)}
     {\prod\Gamma\left(\nu_i\right)}
\int\limits_0^1 \ldots \int\limits_0^1
\frac{\prod \alpha_i^{\nu_i-1} \mbox{d}\alpha_i \;\; 
      \delta\left( \sum\alpha_i -1 \right)}
     {\left[ 
\begin{array}{c} {} \\[-3mm] {\sum\sum} \\[-3mm] {}_{j<l} \end{array} 
\alpha_j \alpha_l k_{jl}^2
             - \sum \alpha_i m_i^2 \right]^{\Sigma\nu_i - n/2} } .
\end{equation}
As we have already mentioned (\ref{homo}), the integrand can be written 
in a homogeneous form:
\begin{equation}
\label{Fp2}
J^{(N)}\left(n; \nu_1, \!\ldots\! , \nu_N \right)
= \mbox{i}^{1\!-2\Sigma\nu_i} \pi^{n/2}
\frac{\Gamma\left(\sum\nu_i \!-\! {\textstyle{n\over2}} \right)}
     {\prod\Gamma\left(\nu_i\right)} \!
\int\limits_0^1 \!\ldots\! \int\limits_0^1 \!
\frac{\prod \alpha_i^{\nu_i-1} \mbox{d}\alpha_i \;\;
      \delta\left( \sum\alpha_i -1 \right)}
     {\left[ \sum \alpha_i^2 m_i^2 \!+\!
2\!\begin{array}{c} {} \\[-3mm] {\sum\sum} \\[-3mm] {}_{j<l} 
    \end{array}
\alpha_j \alpha_l m_j m_l c_{jl} \right]^{\Sigma\nu_i \!-\! n/2} } 
\end{equation}
where the ``cosines'' $c_{jl}$ are defined in (\ref{def_c}).
Note that the limits of integration in eqs.~(\ref{Fp1})--(\ref{Fp2})
can be extended from $(0,1)$ to $(0,\infty)$, since the actual region
of integration is defined by the $\delta$ function.
It corresponds to a part of an $(N-1)$-dimensional hyperplane
$\sum\alpha_i=1$ which is cut out by the conditions 
$\alpha_i\geq 0 \;\; (i=1, \ldots, N)$.

Now, let us consider the integral (\ref{Fp2}), and let us use 
a rescaling which is similar to one used in \cite{KSW2,'tHV-79,DNS}
(see also in \cite{Scharf} where the transformations of a general
form have been discussed). Let us rescale 
$\alpha_i = m_i^{-1}\alpha'_i$, so that the $\delta$ function becomes
\begin{equation}
\delta\left( \sum\frac{\alpha'_i}{m_i} - 1 \right) .
\end{equation}
To restore the argument of the $\delta$ function in its original form,
let us substitute
\begin{equation}
\label{def_F}
\alpha'_i = {\cal{F}}(\alpha''_1, \ldots , \alpha''_N) \alpha''_i,
\hspace{13mm} \mbox{with} \hspace{10mm}
{\cal{F}}(\alpha_1, \ldots , \alpha_N)
= \frac{\sum\alpha_i}{\sum\frac{\alpha_i}{m_i}} .
\end{equation}
Note that the Jacobian of this substitution is ${\cal{F}}^N$
(cf. Appendix~B).

Suppressing the primes, we arrive at the following representation:
\begin{equation}
\label{Fp3}
J^{(N)}\left(n; \nu_1, \ldots , \nu_N \right) \!
= \mbox{i}^{1-2\Sigma\nu_i} \pi^{n/2}
\frac{\Gamma\left(\sum\nu_i \!-\! {\textstyle{n\over2}} \right)}
     {\prod\Gamma\left(\nu_i\right)}
\; \frac{1}{\prod m_i^{\nu_i}}
\int\limits_0^1 \!\ldots\! \int\limits_0^1
\frac{\prod \alpha_i^{\nu_i-1} \mbox{d}\alpha_i \;\;
      \delta\left( \sum\alpha_i -1 \right)}
     { \left( \sum\frac{\alpha_i}{m_i}\right)^{n-\Sigma\nu_i}
\left( \alpha^T \|c\| \alpha \right)^{\Sigma\nu_i - n/2} } ,
\end{equation}
where we use matrix notation (\ref{c-matrix}),
\begin{equation}
\label{matrix_not}
 \alpha^T \|c\| \alpha \equiv
\sum_{j=1}^N \sum_{l=1}^N c_{jl} \alpha_j \alpha_l
=  \sum \alpha_i^2  +
2\!\begin{array}{c} {} \\[-3mm] {\sum\sum} \\[-3mm] {}_{j<l}
\end{array}
\alpha_j \alpha_l c_{jl} .
\end{equation} 

Next, let us change the variables from $\alpha$ to $\alpha'$ via
\begin{equation}
\label{sub3}
\alpha_i={\cal{G}}(\alpha'_1, \ldots , \alpha'_N) \alpha'_i,
\hspace{10mm} \mbox{with} \hspace{5mm}
{\cal{G}}(\alpha'_1, \ldots , \alpha'_N)
=\frac{\sum(\alpha'_i)^2+
2\!\begin{array}{c} {} \\[-3mm] {\sum\sum} \\[-3mm] {}_{j<l} \end{array}
          \! \alpha'_j \alpha'_l c_{jl}}
      {\sum\alpha'_i} .
\end{equation}
Here, the Jacobian of the substitution is $2{\cal{G}}^N$ (see
Appendix~B). Effectively,
we put the quadratic form into the argument of the $\delta$ function.
However, the linear denominator (coming from the ${\cal{F}}$ function)
survives and we arrive at
\begin{equation}
\label{Fp4}
J^{(N)}\left(n; \nu_1, \ldots , \nu_N \right)
= 2 \mbox{i}^{1-2\Sigma\nu_i} \pi^{n/2}
\frac{\Gamma\left(\sum\nu_i \!-\! {\textstyle{n\over2}} \right)}
     {\prod\Gamma\left(\nu_i\right)}
\; \frac{1}{\prod m_i^{\nu_i}}
\int\limits_0^{\infty} \!\ldots\! \int\limits_0^{\infty}
\frac{\prod \alpha_i^{\nu_i-1} \mbox{d}\alpha_i}
     {\left( \sum\frac{\alpha_i}{m_i} \right)^{n-\Sigma\nu_i}}
\delta\left(
\alpha^T \|c\| \alpha
\!-\! 1 \right) ,
\end{equation}
where we continue to use the matrix notation (\ref{matrix_not}).

\subsection{The case $\sum\nu_i=n$}

In the particular case $\sum\nu_i = n$ the linear denominator
disappears and eq.~(\ref{Fp3}) gives
\begin{equation}
\label{uniq1}
\left.  J^{(N)}\left(n; \nu_1, \ldots , \nu_N \right)
\right|_{\Sigma\nu_i=n}
= \mbox{i}^{1-2\Sigma\nu_i} \pi^{n/2}
\frac{\Gamma\left({\textstyle{n\over2}} \right)}
     {\prod\Gamma\left(\nu_i\right)}
\; \frac{1}{\prod m_i^{\nu_i}}
\int\limits_0^1 \!\ldots \!\int\limits_0^1
\frac{\prod \alpha_i^{\nu_i-1} \mbox{d}\alpha_i \;\;
      \delta\left( \sum\alpha_i -1 \right)}
{\left( \alpha^T \|c\| \alpha \right)^{\Sigma\nu_i - n/2} } .
\end{equation}
Here all dependence on the external momenta and masses 
in the integral is through $c_{jl}$, eq.~(\ref{def_c}).

Since the final integral (\ref{uniq1}) is the same as in the
equal-mass case,
we can formulate the following statement:
Let $\sum\nu_i=n$, and let the result for the Feynman integral 
(\ref{Fp1}) with equal
masses ($m_i=m$, $i=1,\ldots,N$) be represented by 
a dimensionless function $\Psi$
depending on the quantities 
$\overline{\overline{c}}_{jl}\equiv 1 - k_{jl}^2/(2m^2)$
(corresponding
to eq.~(\ref{def_c}) in the equal-mass case) as
\begin{equation}
\label{eqm1}
\left. \; J^{(N)}\left(n; \nu_1, \ldots , \nu_N \right)
\right|_{\Sigma\nu_i=n, \;\; m_i=m}
=  \frac{1}{m^{\Sigma\nu_i}} \; 
\Psi\left(\left\{ \overline{\overline{c}}_{jl} \right\}\right) .
\end{equation}
Then the result for the corresponding integral with different
masses $m_i$ can be expressed in terms of {\em the same} 
function $\Psi$ as
\begin{equation}
\label{eqm2}
\left. \; J^{(N)}\left(n; \nu_1, \ldots , \nu_N \right)
\right|_{\Sigma\nu_i=n}
=  \frac{1}{\prod m_i^{\nu_i}} \;
\Psi\left(\left\{ c_{jl} \right\}\right) , 
\end{equation}
where $c_{jl}$ are defined in eq.~(\ref{def_c}).

Simplification of the integrals $J^{(N)}(n;\nu_1,\ldots,\nu_N)$
when $\sum\nu_i=n$ is not surprising;
to some extent, this may be considered a generalization
of the so-called ``uniqueness'' formula for massless
triangle diagrams \cite{uniq} to the case of massive $N$-point
integrals. 

Whenever $\sum\nu_i=n$ the denominator of the integrand 
of (\ref{Fp4}) disappears, and the result is
\begin{equation}
\label{uniq2}
\left. \! J^{(N)}\left(n; \nu_1, \!\ldots\! , \nu_N
\right)\right|_{\Sigma\nu_i=n} \!
= 2 \mbox{i}^{1-2\Sigma\nu_i} \pi^{n/2}
\frac{\Gamma\left( {\textstyle{n\over2}} \right)}
     {\prod\Gamma\left(\nu_i\right)}
\; \frac{1}{\prod m_i^{\nu_i}}
\int\limits_0^{\infty} \!\ldots\! \int\limits_0^{\infty}
\prod \alpha_i^{\nu_i-1} \mbox{d}\alpha_i
\; \delta\left( 
\alpha^T \|c\| \alpha
\!-\! 1 \right) .
\end{equation}
In particular, when all $\nu_i=1$ ($n=N$), our integrand is just a
$\delta$ function,
\begin{equation}
\label{uniq3}
J^{(N)}\left(N; 1, \ldots , 1 \right)
= 2\; \mbox{i}^{1-2N} \pi^{N/2}
\frac{\Gamma\left( {\textstyle{N\over2}} \right)}
     {\prod m_i}
\int\limits_0^{\infty} \ldots \int\limits_0^{\infty}
\prod\mbox{d}\alpha_i
\; \delta\left( 
\alpha^T \|c\| \alpha
-1 \right) .
\end{equation}

Consider also a very special situation, when all non-diagonal $c_{jl}$
vanish (i.e. $c_{jl}=\delta_{jl}$).
Physically, this corresponds to the case where all external momenta 
squared take the mean values between the threshold and the 
pseudo-threshold, $k_{jl}^2=m_j^2+m_l^2$ (for $j<l$). In this case,
all the angles $\tau_{jl}$ ($j\neq l$) are equal to 
${\textstyle{1\over2}}\pi$.
Employing the representation (\ref{uniq2}) we arrive at a
very simple result for this special case,
\begin{equation}
\label{special2}
\left. \; J^{(N)}\left(n; \nu_1, \ldots , \nu_N \right)
\right|_{\begin{array}{l} {}_{\!\! \Sigma\nu_i=n} \\ 
                          {}^{\!\! c_{jl}=\delta_{jl}}
         \end{array}}
= \mbox{i}^{1-2\Sigma\nu_i} \; \pi^{n/2} \;
\frac{\prod\Gamma\left({\textstyle{1\over2}} \nu_i \right)}
     {2^{N-1} \prod\Gamma\left(\nu_i\right)}
\; \frac{1}{\prod m_i^{\nu_i}} .
\end{equation}

Imagine the $N$-dimensional Euclidean space of the parameters 
$\alpha_i$. The quadratic argument of the $\delta$ function in
(\ref{uniq3}) defines a 
$(N-1)$-dimensional hypersurface. Provided that all eigenvalues of
this quadratic form are positive, this is an $N$-dimensional  
ellipsoid. Otherwise, this is an  $N$-dimensional hyperboloid.
So, the integral (\ref{uniq3}) is nothing but the measure 
(i.e. the content) of
the part of this hypersurface which belongs to the region where 
all $\alpha$'s are non-negative. This region is a $2^N$-th part
of the whole $N$-dimensional space.

To calculate the content of this part of the hypersurface, 
it is reasonable to 
diagonalize the quadratic form. Suppose there is an $N$-dimensional 
rotation transforming the old coordinates $\alpha_i$ into the new ones,
$\beta_i$, so that 
\begin{equation}
\alpha^T \|c\| \alpha =
\sum \alpha_i^2  +
2\!\begin{array}{c} {} \\[-3mm] {\sum\sum} \\[-3mm] {}_{j<l} \end{array}
\alpha_j \alpha_l c_{jl}
\Rightarrow
\sum \lambda_i \beta_i^2 .
\end{equation}
Obviously, the product of $\lambda$'s must be equal to the
determinant $D^{(N)}$ of the matrix (\ref{c-matrix}) 
corresponding to the quadratic form (\ref{matrix_not}),
\begin{equation}
\label{prod_lambda}
\lambda_1 \ldots \lambda_N = D^{(N)} . 
\end{equation}  
Note that this determinant $D^{(N)}$ is related to the
content $V^{(N)}$ of the basic $N$-dimensional simplex via
eq.~(\ref{V^{(N)}}).

Now, let us assume that all eigenvalues $\lambda_i$ are real and
positive, i.e. the hyper-surface defined by the quadratic form is 
an $N$-dimensional ellipsoid. 
After rotation from $\alpha$'s to $\beta$'s,
we get into the co-ordinate system corresponding to the principal
axes of the ellipsoid. Then, we can rescale 
\begin{equation}
\beta_i=\frac{\gamma_i}{\sqrt{\lambda_i}},
\end{equation}
the Jacobian of this transformation being 
$\left( \lambda_1 \ldots \lambda_N \right)^{-1/2} 
= \left( D^{(N)} \right)^{-1/2}$, according to (\ref{prod_lambda}).
In this way, our $N$-dimensional ellipsoid is transformed 
into a hypersphere.
Now all we need to calculate is the content of a part of this
hypersphere which is cut out (in the space of $\gamma_i$) by the images 
of the hyperfaces restricting the region where all $\alpha_i$ are
positive (in the space of $\alpha_i$). This content, which we shall 
denote by $\Omega^{(N)}$, can be understood as the $N$-dimensional 
solid angle subtended by the above-mentioned hyperfaces. 
In terms of $\Omega^{(N)}$,
the relevant integral (\ref{uniq3}) can be written as
\begin{equation}
\label{uniq4}
J^{(N)}\left(N; 1, \ldots , 1 \right)
= \mbox{i}^{1-2N} \; \pi^{N/2} \;
\frac{\Gamma\left( {\textstyle{N\over2}} \right)}
     {\prod m_i} \;\;
\frac{\Omega^{(N)}}{\sqrt{D^{(N)}}} .
\end{equation}

An interesting fact is that $\Omega^{(N)}$ is also related to the
basic simplex. To see this, let us note that 
effectively (in terms of $\alpha$'s), the transformation
$\alpha_i\rightarrow\beta_i\rightarrow\gamma_i$
is equivalent to using the matrix $\|c\|$ as a new metric.
Say, for a vector $\alpha$ the {\em new} length squared is
\begin{equation}
|\alpha|_c^2 \equiv \alpha^T \|c\| \alpha .
\end{equation}
For unit vectors $e_i^{(\alpha)}$, the {\em new} scalar product is
\begin{equation}
\left(e_j^{(\alpha)}\cdot e_l^{(\alpha)}\right)_c = c_{jl} 
\equiv \cos\tau_{jl} .
\end{equation}
Taking into account that $|e_i^{(\alpha)}|_c^2=1$,
we see that in the space with the $c$-metric (i.e. in the
space of $\gamma_i$) the angle between
$e_j^{(\alpha)}$ and $e_l^{(\alpha)}$ is $\tau_{jl}$. 
Therefore, we get nothing but the $N$-dimensional solid angle at 
the mass meeting point of the basic simplex. It defines the region of 
integration on the hypersurface of the unit hypersphere in the 
$\gamma_i$ coordinate system.
We also note that the cosines $c_{jl}$ are invariant under 
simple rescaling of $\alpha$'s.

It is easy to see that the image of $e_i^{(\alpha)}$ in the 
$\gamma_i$-space is directed along the vector $a_i$. 
Moreover, according to eq.~(\ref{n_0_a}), the image of
the vector with the components $F_i^{(N)}/m_i$ is directed
along the height $H_0$. 

To summarize, the following statement is valid:
The content of the $N$-dimensional solid angle $\Omega^{(N)}$ 
in the space
of $\gamma_i$ is equal to that at the vertex of the basic 
$N$-dimensional simplex
where all mass sides meet. Moreover, the angles between the
corresponding hyperfaces in the space of $\gamma_i$ and those
in the basic simplex are the same. Therefore, the result for
the integral (\ref{uniq3}) can be expressed in terms of the
content of the basic simplex and the content of its $N$-dimensional 
solid angle at the vertex which is common for all mass sides as
\begin{equation}
\label{uniq5}
J^{(N)}\left(N; 1, \ldots , 1 \right)
= \mbox{i}^{1-2N} \; \pi^{N/2} \;
\frac{\Gamma\left( {\textstyle{N\over2}} \right)}
     {N!} \;\;
\frac{\Omega^{(N)}}{V^{(N)}} .
\end{equation}

Looking at eq.~(\ref{uniq5}), we see that $\Omega^{(N)}$ is 
indeed the only thing which is to be calculated, since $V^{(N)}$ is 
known through eq.~(\ref{V^{(N)}}). In the following sections, we shall
consider the lowest examples to understand how to calculate 
$\Omega^{(N)}$. 

Nevertheless, the above discussion leads us to
the following significant statement:   
The content $\Omega^{(N)}$ is nothing
but the content of a {\em non-Euclidean} $(N-1)$-dimensional simplex
calculated in the spherical (or hyperbolic,
depending on the signature of the eigenvalues $\lambda_i$) space
of constant curvature.
The sides of this non-Euclidean simplex are equal to
the angles $\tau_{jl}$. Therefore, the problem of calculating
Feynman integrals is intimately connected with the problem of
calculating the content of a simplex in non-Euclidean geometry.
   
\subsection{The general case}

To understand how to deal with the general case, when 
$\Sigma\nu_i\neq n$, we need some modification of the above
transformations. First of all, we note that eqs.~(\ref{linear})
can be rephrased as
\begin{equation}
\label{linear2}
\sum_{l=1}^{N} 
\left( \sqrt{F_j^{(N)}} \; c_{jl} \; \sqrt{F_l^{(N)}} \right)
\frac{\sqrt{F_l^{(N)}}}{m_l}
= D^{(N)} \; \frac{\sqrt{F_j^{(N)}}}{m_j} .
\end{equation}
In other words, the vector with components
\begin{equation}
\label{f_i}
f_i = \frac{\sqrt{F_i^{(N)}}}{m_i}
\end{equation}
is an eigenvector of the matrix $\|C\|\equiv\|C_{jl}\|$ with the
elements
\begin{equation}
\label{C_{jl}}
C_{jl}
= \left( \sqrt{F_j^{(N)}} \; c_{jl} \; \sqrt{F_l^{(N)}} \right) ,
\end{equation} 
and the corresponding eigenvalue is equal to $D^{(N)}$. Note that
\begin{equation}
f^2 \equiv (f\cdot f) 
= \sum_{i=1}^N \frac{F_i^{(N)}}{m_i^2}
= \frac {\Lambda^{(N)}} 
        {\prod_{i=1}^N m_i^2} \; ,
\end{equation}
\begin{equation}
\det\|C\| = D^{(N)} \; \prod_{j=1}^N F_j^{(N)} .
\end{equation}
 
The idea is that it may be more convenient to use the quadratic
form defined by the matrix $\|C_{jl}\|$, rather than $\|c_{jl}\|$,
since we know one of its eigenvectors with the corresponding
eigenvalue.

After using the masses $m_i$ to rescale the Feynman parameters 
$\alpha_i$ and ``hiding'' the quadratic form into the argument 
of the delta function, we obtained the representation
(\ref{Fp4}).
Analogously, we can use $1/f_i=m_i/\sqrt{F_i^{(N)}}$, rather than $m_i$,
to rescale the original $\alpha$'s. The result is
\begin{eqnarray}
\label{f1}
J^{(N)}\left(n;\nu_1,\ldots,\nu_N\right)
\hspace{110mm}
\nonumber \\
= 2 \mbox{i}^{1-2\Sigma\nu_i} \pi^{n/2}
\frac{\Gamma\left(\sum\nu_i \!-\! \frac{n}{2}\right)}
     {\prod \Gamma\left(\nu_i\right) } \;
\left(\prod f_i^{\nu_i} \right)
\int\limits_0^{\infty} \! \ldots \! \int\limits_0^{\infty} 
\frac{\prod \alpha_i^{\nu_i-1} \mbox{d}\alpha_i}
     {\left( \sum \alpha_i f_i \right)^{n-\Sigma\nu_i}} \;
\delta\left( \alpha^T \|C\| \alpha \!-\! 1 \right)\! ,
\end{eqnarray}
where notation (\ref{matrix_not}) has been used again. 
A nice feature of this representation is that in the denominator
we have got the contraction (scalar product) of $\alpha$ and
the known eigenvector of the matrix $\|C\|$ defining 
the quadratic form (see eqs.~(\ref{linear2})--(\ref{C_{jl}})).

If $\nu_1=\ldots =\nu_N=1$, we get
\begin{equation}
\label{f2}
J^{(N)}\left(n; 1,\ldots, 1\right)
= 2 \; \mbox{i}^{1-2N} \; \pi^{n/2} \;
\Gamma\left(N \!-\! {\textstyle{n\over2}}\right) \;
\left(\prod f_i \right) \;
\int\limits_0^{\infty} \! \ldots \! \int\limits_0^{\infty} 
\frac{\prod \mbox{d}\alpha_i}
     {\left( \sum \alpha_i f_i \right)^{n-N}} \;
\delta\left( \alpha^T \|C\| \alpha - 1 \right) .
\end{equation}
Now, let us use the same transformations of variables as earlier.
Namely, let us first rotate from $\alpha_i$ to the variables $\beta_i$
such that the quadratic form becomes diagonal,
\begin{equation}
\alpha^T \|C\| \alpha = \sum_{i=1}^N \lambda_i \beta_i^2 .
\end{equation}
One of these $\beta$'s, say $\beta_N$, is directed along the
eigenvector $f$. Therefore, $\lambda_N=D^{(N)}$ and
$\prod_{i=1}^{N-1} \lambda_i = \prod_{j=1}^N F_j^{(N)}$.
Moreover, the denominator is also proportional to $\beta_N$.
Finally, we rescale $\beta_i=\gamma_i/\sqrt{\lambda_i}$, and
the denominator becomes
\begin{equation}
\left( \sum \alpha_i f_i \right) \Rightarrow
\frac{1}{\prod m_i} \sqrt{\frac{\Lambda^{(N)}}{D^{(N)}}} \; \gamma_N
= \frac{1}{m_0} \; \gamma_N ,
\end{equation}
where $m_0$, defined in eq.~(\ref{m_0}),
is the distance between
the mass meeting point and the momentum hyperface
(i.e. the length of the height $H_0$).

Now, the transformation
$\alpha_i\rightarrow\beta_i\rightarrow\gamma_i$
is equivalent to using the matrix $\|C\|$ as a new metric.
For a vector $\alpha$ the {\em new} length squared is
\begin{equation}
|\alpha|_C^2 \equiv \alpha^T \|C\| \alpha .
\end{equation}
For unit vectors $e_i^{(\alpha)}$, the {\em new} scalar product is
\begin{equation}
\left(e_j^{(\alpha)}\cdot e_l^{(\alpha)}\right)_C = C_{jl} ,
\end{equation}
and the corresponding cosine of the angle is\footnote{Note that
in the $\gamma_i$-space with the $C$-metric the $e_i^{(\alpha)}$
vectors are not unit, $|e_i^{(\alpha)}|_C^2=C_{ii}=F_i^{(N)}$.}
\begin{equation}
\frac{\left(e_j^{(\alpha)}\cdot e_l^{(\alpha)}\right)_C} 
     {|e_j^{(\alpha)}|_C \; |e_l^{(\alpha)}|_C}
= \frac{C_{jl}}{\sqrt{C_{jj}\; C_{ll}}} = c_{jl} \equiv \cos\tau_{jl} .
\end{equation}
Again, as in the case of the $c$-metric, we get the same
$N$-dimensional solid angle as that occurring at the mass meeting 
point of the basic simplex. 

But we can go further. In particular, we can comprehend
what is the image of the vector $f$.
Consider the scalar products
\begin{eqnarray}
\left( e_i^{(\alpha)} \cdot f \right)_C
= \left( e_i^{(\alpha)} \cdot \|C\| f \right)
= D^{(N)} \left( e_i^{(\alpha)} \cdot f \right)
= D^{(N)} f_i , 
\\
\left( f \cdot f \right)_C 
= \left( f \cdot \|C\| f \right)
= D^{(N)} \left( f \cdot f \right)
= \frac{D^{(N)} \; \Lambda^{(N)}}{\prod_{j=1}^N m_j^2} .
\hspace{20mm}
\end{eqnarray}
Hence, the cosine of the angle between the images of 
$f$ and $e_i^{(\alpha)}$ is (cf. eq.~(\ref{c0j}))
\begin{equation}
\frac{\left( e_i^{(\alpha)} \cdot f \right)_C}
     {\sqrt{\left(e_i^{(\alpha)} \cdot e_i^{(\alpha)}\right)_C \;
            \left(f \cdot f \right)_C}}
= \sqrt{\frac{D^{(N)}}{\Lambda^{(N)}}} \; \prod_{j\neq i}^{} m_j
= \frac{m_0}{m_i} = \cos\tau_{0i} \equiv c_{0i} .
\end{equation}
Therefore, in the basic simplex, with the mass sides $m_i$, etc.,
the image of the vector $f$ is directed along the height $H_0$~!

In terms of $\gamma$'s, the integral becomes
\begin{equation}
\label{gen}
J^{(N)}(n; 1, \ldots, 1) = 2\mbox{i}^{1-2N} \pi^{n/2} \;
\Gamma\left(N-{\textstyle{n\over2}}\right) \;
\frac{m_0^{n-N-1}}{\sqrt{\Lambda^{(N)}}} \;
\begin{array}{c} {} \\ \int\limits \ldots \int\limits \\ 
                 {}^{\Omega^{(N)}} \end{array}
\frac{\prod \mbox{d}\gamma_i}{\gamma_N^{n-N}} \;
\delta\left( \sum\gamma_i^2 -1\right) ,
\end{equation}
where the integration goes over the interior of the $N$-dimensional
solid angle
$\Omega^{(N)}$ of the basic simplex.
If we define the angle between the ``running'' unit vector and the
$N$-th axis as $\theta$, the above formula means that we should
integrate over the hypersurface of the unit hypersphere with
the ``weight'' $(\cos\theta)^{N-n}$, within the limits set by
$\Omega^{(N)}$,
\begin{equation}
\label{gen2}
J^{(N)}(n; 1, \ldots, 1) = \mbox{i}^{1-2N} \pi^{n/2} \;
\Gamma\left(N-{\textstyle{n\over2}}\right) \;
\frac{m_0^{n-N}}{\sqrt{D^{(N)}}\; \prod m_i} \;
\Omega^{(N;n)} ,
\end{equation}
with 
\begin{equation}
\label{OmegaNn}
\Omega^{(N;n)} \equiv
\begin{array}{c} {} \\ \int\limits \ldots \int\limits \\
                 {}^{\Omega^{(N)}} \end{array}
\frac{\mbox{d}\Omega_N}{\cos^{n-N}\theta} .
\end{equation}
Obviously, $\Omega^{(N;N)}=\Omega^{(N)}$. Using eq.~(\ref{V^{(N)}}),
the formula (\ref{gen2}) can be exhibited in the following
form
\begin{equation}
\label{gen3}
J^{(N)}(n; 1, \ldots, 1) = \mbox{i}^{1-2N} \pi^{n/2} \;
\Gamma\left(N-{\textstyle{n\over2}}\right) \;
\frac{m_0^{n-N} \; \Omega^{(N;n)}}{N! \; V^{(N)}} .
\end{equation}  

\subsection{Splitting the $N$-dimensional solid angle}

Now, let us again use the height $H_0$ to split the basic
$N$-dimensional simplex into $N$ rectangular ones, each time 
replacing one of the mass sides, $m_i$, by $m_0=|H_0|$.
The content of the $i$-th rectangular simplex $V_i^{(N)}$
is given by eq.~(\ref{V_i^{(N)}}). Let us denote the corresponding
$N$-dimensional solid angle as $\Omega_i^{(N)}$, so that
\begin{equation}
\label{Omega_j^{(N;n)}}
\Omega_i^{(N;n)} =
\begin{array}{c} {} \\ \int\limits \ldots \int\limits \\
                 {}^{\Omega_i^{(N)}} \end{array}
\frac{\mbox{d}\Omega_N}{\cos^{n-N}\theta}, \hspace{8mm} 
\Omega^{(N;n)} = 
\sum_{i=1}^N \Omega_i^{(N;n)}.
\end{equation}
Substituting  (\ref{Omega_j^{(N;n)}}) into eq.~(\ref{gen3}), we get
\begin{equation}
\label{split1_JN}
J^{(N)}(n;1, \ldots, 1) = \sum_{i=1}^N \frac{V_i^{(N)}}{V^{(N)}} \;
J_i^{(N)}(n;1, \ldots, 1) , 
\end{equation}
where $J_i^{(N)}$ are the integrals corresponding to the 
rectangular simplices. Specifically, in the integral $J_i^{(N)}$
the internal masses are 
\begin{equation}
\label{mas_i}
m_1, \ldots, m_{i-1}, m_0, m_{i+1}, \ldots, m_N ,
\end{equation}
while the momenta invariants are
\begin{equation}
\label{mom_i}
\left\{ \begin{array}{l} 
k_{jl}^2 , \hspace{5mm} \mbox{if} \;\; 
    j\neq i \hspace{3mm} \mbox{and} \hspace{3mm} l\neq i \\
m_l^2 - m_0^2 , \hspace{5mm} \mbox{if} \;\; j=i \\
m_j^2 - m_0^2 , \hspace{5mm} \mbox{if} \;\; l=i
\end{array} . \right. 
\end{equation}
Using (\ref{V_i^{(N)}}), we arrive at
\begin{equation}
\label{split2_JN}
J^{(N)}(n;1, \ldots, 1) = \frac{1}{\Lambda^{(N)}}\; 
\left(\prod m_i^2 \right) \;
\sum_{i=1}^N \frac{1}{m_i^2} \; F_i^{(N)} \;
J_i^{(N)}(n;1, \ldots, 1) .
\end{equation}
 
We are now ready to tackle some specific examples using the 
geometrical approach.

\section{Two-point function}
\setcounter{equation}{0}

Here, the two-dimensional basic simplex
is a triangle and the angle $\tau_{12}$ lies between the two mass sides
(see  Fig.~4). We obtain
\begin{eqnarray}
\label{two-point}
D^{(2)}= 1-c_{12}^2 = \sin^2\tau_{12}, \hspace{5mm}
V^{(2)}={\textstyle{1\over2}} \;m_1 m_2 \; \sin\tau_{12}, \hspace{5mm}
\Omega^{(2)}=\tau_{12} ,
\\
m_0 = m_1 m_2 \sqrt{\frac{D^{(2)}}{\Lambda^{(2)}}},
\hspace{5mm}
\cos\tau_{0i} = \frac{m_0}{m_i}, \hspace{5mm}
\tau_{01}+\tau_{02} = \tau_{12}, \hspace{5mm}
\Lambda^{(2)}=k_{12}^2 .
\end{eqnarray}
In two dimensions, from (\ref{uniq5}) we obtain the well-known result
\begin{equation}
J^{(2)}(2;1,1) = \frac{\mbox{i}\pi}{m_1 m_2} \;
                 \frac{\tau_{12}}{\sin\tau_{12}} ,
\end{equation}
while in three dimensions, eqs.~(\ref{gen3})--(\ref{Omega_j^{(N;n)}})
yield
\begin{equation}
J^{(2)}(3;1,1) = \mbox{i} \pi^2 \frac{1}{\sqrt{\Lambda^{(2)}}} \;
\left\{ \Omega_1^{(2;3)} + \Omega_2^{(2;3)} \right\} ,
\end{equation}
with
\begin{equation}
\Omega_i^{(2;3)} =
\int\limits_0^{\tau_{0i}} \frac{\mbox{d}\theta}{\cos\theta}
=  \ln\left(\frac{1+\sin\tau_{0i}}{1-\sin\tau_{0i}} \right) . 
\end{equation}
Combining the logarithms, we get
\begin{equation}
J^{(2)}(3;1,1) = \frac{\mbox{i}\pi^2}{\sqrt{k_{12}^2}}
\ln\left(\frac{m_1+m_2+\sqrt{k_{12}^2}}{m_1+m_2-\sqrt{k_{12}^2}} 
\right) .
\end{equation}
In the Euclidean region ($k_{12}^2<0$) the logarithm
gives $\arctan(\sqrt{-k_{12}^2}/(m_1+m_2))$, and the result
coincides with those presented in \cite{Nickel,Rajantie}.

In four dimensions, the $\Gamma$ function in front of the
integral (\ref{gen2}) becomes singular. Introducing dimensional
regularization \cite{dimreg} to circumvent that difficulty, we get
\begin{equation}
\label{J^2(4,1,1)}
J^{(2)}(4-2\varepsilon;1,1)
=\mbox{i}\pi^{2-\varepsilon}\Gamma(\varepsilon)
\frac{m_0^{1-2\varepsilon}}{\sqrt{\Lambda^{(2)}}} \; 
\left\{ \Omega_1^{(2; 4-2\varepsilon)} 
+ \Omega_2^{(2; 4-2\varepsilon)} \right\} ,
\end{equation}
with
\begin{equation}
\label{Omega_i^(2;4-2ep)}
\Omega_i^{(2; 4-2\varepsilon)} =
\int\limits_0^{\tau_{0i}} 
\frac{\mbox{d}\theta}{\cos^{2-2\varepsilon}\theta} .
\end{equation}
Expanding the integrand in $\varepsilon$ and taking into account that
\begin{equation}
\int\limits_0^{\tau} \frac{\mbox{d}\theta}{\cos^2\theta}
\ln(\cos\theta)
= \tan\tau \ln(\cos\tau)+\tan\tau -\tau ,
\end{equation}
we reach the well-known result (cf., e.g., in \cite{'tHV-79,BDS})
\begin{eqnarray}
J^{(2)}(4-2\varepsilon;1,1)= \mbox{i} 
\pi^{2-\varepsilon} \Gamma(1+\varepsilon)
\hspace{70mm}
\nonumber \\
\times
\left\{
\frac{1}{\varepsilon}+2-\ln m_1 -\ln m_2 
+ \frac{m_1^2-m_2^2}{k_{12}^2} \ln\frac{m_2}{m_1}
- \frac{2 m_1 m_2}{k_{12}^2} \tau_{12} \sin\tau_{12} 
\right\} + {\cal{O}(\varepsilon)} .
\end{eqnarray}

Furthermore, the representation 
(\ref{J^2(4,1,1)})--(\ref{Omega_i^(2;4-2ep)}) makes it possible
to construct next terms of the expansion in $\varepsilon$. For example, 
the $\varepsilon$ term requires just\footnote{The Clausen function 
$\mbox{Cl}_2(\theta)$ is defined in Appendix~A.}
\begin{equation}
\int\limits_0^{\tau} \frac{\mbox{d}\theta}{\cos^2\theta}
\ln^2(\cos\theta) = 
\tan\tau \left(\ln^2(\cos\tau)+2\ln(\cos\tau)+2 \right)
-2\tau (1-\ln 2) - \mbox{Cl}_2\left(\pi-2\tau\right) 
\end{equation}
(the corresponding result was obtained in \cite{NMB}).
Moreover, the result for an arbitrary space-time dimension 
or $\varepsilon$ 
can be obtained in terms of the Gauss hypergeometric function
(cf. in \cite{BD-TMF,BDS}), using
\begin{equation}
\Omega_i^{(2; 4-2\varepsilon)} =
\int\limits_0^{\tau_{0i}} 
\frac{\mbox{d}\theta}{\cos^{2-2\varepsilon}\theta}
= 2 \tan\tau_{0i} \;\; 
_2F_1\left( \left. 
\begin{array}{c} 1/2, \; \varepsilon \\ 3/2 \end{array} 
\right| -\tan^2\tau_{0i} \right) .
\end{equation}
Using formulae of analytic continuation of $_2F_1$ function, one
can establish a connection with the result presented in
eq.~(A.7) of \cite{BDS}. 

\section{Three-point function}
\setcounter{equation}{0}

\subsection{Geometrical picture and the three-dimensional case}

Here the three-dimensional basic simplex is a tetrahedron with three  
mass sides (the angles between these mass sides are $\tau_{12},
\tau_{13}$ and $\tau_{23}$) and three momentum sides.
It is shown in Fig.~5a.
The volume of this tetrahedron is defined as
\begin{equation}
V^{(3)}={\textstyle{1\over6}} \; m_1 m_2 m_3 \; \sqrt{D^{(3)}} ,
\end{equation}
where (cf. eqs.~(\ref{c-matrix})--(\ref{D(N)}))
\begin{equation}
\label{D(3)}
D^{(3)} \equiv
\left|
\begin{array}{c} 
{ \; 1\;\;  \;\; c_{12}\; \;\; c_{13}\;  } \\
{ c_{12}\;  \;\; \; 1\;\; \;\; c_{23}\;  } \\
{ c_{13}\;  \;\; c_{23}\; \;\; \; 1\;\;  } \\
\end{array}
\right|
=1-c_{12}^2-c_{13}^2-c_{23}^2+2c_{12} c_{13} c_{23} .
\end{equation}
Furthermore, $\Omega^{(3)}$ is the usual solid angle at the vertex 
derived by the mass sides (cf. Fig.~5b). Its value can be defined 
as the area of 
a part of the unit sphere cut out by the three planar faces adjacent 
to the vertex; in other words, this is the area of a spherical triangle
corresponding to this section. The sides of this spherical triangle   
are obviously equal to the angles $\tau_{12}, \tau_{13}$ and $\tau_{23}$
while its angles, $\psi_{12}, \psi_{13}$ and $\psi_{23}$, are equal
to those between the plane faces (we define $\psi_{12}$ 
as the angle between the sides $\tau_{13}$ and $\tau_{23}$, etc.).
This definition is in agreement with eq.~(\ref{psi_bar}).
Using eq.~(\ref{cos_psi})
or the well-known formulae of the spherical trigonometry,
it is easy to show that
\begin{equation}
\label{phi's}
\cos\psi_{12}
=\frac{\cos\tau_{12}-\cos\tau_{13} \cos\tau_{23}}
                   {\sin\tau_{13} \sin\tau_{23}} ,
\hspace{5mm}
\sin\psi_{12}=\frac{\sqrt{D^{(3)}}}{\sin\tau_{13}
\sin\tau_{23}},
\end{equation}
and analogous expressions for $\psi_{13}$ 
and $\psi_{23}$.
The solid angle corresponding to spherical triangle can be obtained via
\begin{equation}
\label{solid_angle}
\Omega^{(3)}=\psi_{12}+\psi_{13}+\psi_{23}-\pi .
\end{equation}
Using eqs.~(\ref{phi's}) and (\ref{solid_angle}), it is straightforward
to show that
\begin{eqnarray}
\label{Omega^{(3)}}
\Omega^{(3)} &=& 
2 \arccos\left(\frac{1+c_{12}+c_{13}+c_{23}}
                    {\sqrt{2(1+c_{12})(1+c_{13})(1+c_{23})}}\right)
\nonumber \\
         &=& 2 \arcsin\left(\sqrt{\frac{D^{(3)}}
            {2(1+c_{12})(1+c_{13})(1+c_{23})}}\right)
\nonumber \\
         &=& 2 \arctan
\left(\frac{\sqrt{D^{(3)}}}{1+c_{12}+c_{13}+c_{23}}\right) .
\end{eqnarray}
Finally, the result
\begin{equation}
\label{J(3)}
J^{(3)}(3;1,1,1) = -\frac{\mbox{i} \pi^2}{2 m_1 m_2 m_3} \;
                    \frac{\Omega^{(3)}}{\sqrt{D^{(3)}}} ,
\end{equation}
with $\Omega^{(3)}$ defined by (\ref{Omega^{(3)}}), corresponds to one
obtained in \cite{Nickel} in a different way.
Note that here we have derived it purely geometrically. 
   
However, eq.~(\ref{solid_angle}) cannot be easily generalized to the 
four-dimensional case.
This is why it is instructive to reproduce
the three-dimensional result (\ref{Omega^{(3)}}) in a different manner,
splitting the basic tetrahedron into rectangular ones
(cf. eq.~(\ref{split1_JN})).

Consider a spherical triangle 123 on the unit sphere,
cut out by the solid angle of the basic tetrahedron in the
$\gamma$ space, as illustrated in Fig.~6.
The points 1,2,3 correspond to the intersections of the
mass sides of the basic tetrahedron and the unit sphere.
The sides of the triangle are equal to the angles $\tau_{12}$, 
$\tau_{23}$ and $\tau_{31}$. 
There is another point 0, also on the unit sphere,
corresponding to the intersection of the height $H_0$
of the tetrahedron and the unit sphere. The point 0 is connected
with each of the points 1,2,3. 
The corresponding sides 01, 02 and 03 are
equal to $\tau_{01}$, $\tau_{02}$ and $\tau_{03}$, respectively, so that
\begin{equation}
\cos\tau_{0i} = \frac{m_0}{m_i}, \hspace{10mm}
m_0
=m_1 m_2 m_3 \sqrt{\frac{D^{(3)}}{\Lambda^{(3)}}} ,
\end{equation}
with (cf. eq.~(\ref{Lambda(N)}))
\begin{equation}
\Lambda^{(3)}=-{\textstyle{1\over4}}
\left[ (k_{12}^2)^2 + (k_{13}^2)^2 + (k_{23}^2)^2
       - 2 k_{12}^2 k_{13}^2 - 2 k_{12}^2 k_{23}^2
       - 2 k_{13}^2 k_{23}^2 
\right]
= -{\textstyle{1\over4}} \lambda\left(k_{12}^2,k_{13}^2,k_{23}^2\right) ,
\end{equation}
where $\lambda(x,y,z)$ is defined in eq.~(\ref{Kallen}) (cf. Fig.~1).
The triangle 123 is thereby split into three triangles 012, 023 and 031.
The angle between 01 and 02 is denoted as $\varphi_{12}$, etc.
Obviously,
\begin{equation}
\varphi_{12}+\varphi_{23}+\varphi_{31}=2\pi .
\end{equation}

For the calculation of three-dimensional triangle integral, it does
not really matter where the point 0 lies, though this becomes essential
for the four-dimensional case. As we know, the $N=n$ integrals
can be reduced to the integrals with equal masses. So, let us start
with the equal mass case. In Fig.~6, this means that 
$\tau_{01}=\tau_{02}=\tau_{03}\equiv\tau_0$. Therefore, the spherical
triangles 012, 023 and 031 are isosceles. Consider one
of these triangles, say 012, as illustrated in Fig.~7.
The sides 01 and 02 are equal to $\tau_0$.
The third side, 12, is equal to $\tau_{12}$. The length of
the perpendicular dropped from the point 0 to the side 12
is denoted as $\eta_{12}$. This perpendicular 
intersects the side 12 in a point $T_{12}$ which 
in the equal-mass case
divides the side 12 into two equal parts (each of them equals
$\tau_{12}/2$). It also divides the angle $\varphi_{12}$ into
two equal parts.
The angles at the vertices 1 and 2 
(which are also equal in the equal-mass case)
are denoted as $\kappa_{12}$. 
The lengths of two of the sides of 
a smaller isosceles triangle $01_{\xi}2_{\xi}$, 
$01_{\xi}$ and $02_{\xi}$, are 
$\tau_0\xi \;\; (0\leq \xi\leq 1)$. The third side, $1_{\xi}2_{\xi}$, 
is parallel (in the spherical sense) to the side 12 and its length is
denoted as $\tau(\xi)$; obviously, $\tau(0)=0$ and $\tau(1)=\tau_{12}$.
The perpendicular $0T_{12}$ also splits it into two equal parts
of the length $\tau(\xi)/2$. The length of the part of the perpendicular
within the triangle $01_{\xi}2_{\xi}$ is denoted as $\eta(\xi)$,
whereas the angles at the vertices $1_{\xi}$
and $2_{\xi}$ are $\kappa(\xi)$; obviously, $\eta(0)=0, \; 
\eta(1)=\eta_{12}, \; \kappa(0)=(\pi-\varphi_{12})/2, \; 
\kappa(1)=\kappa_{12}$.

The main idea of introducing an auxiliary variable $\xi$ is to
simplify the limits of integration. We define the usual spherical
integration angle $\theta$ in a way that it is the angle between
the direction of $H_0$ and the ``running'' point. For the
triangle 012 presented in Fig.~7, let us define that $\varphi=0$
corresponds to the direction of the perpendicular $0T_{12}$.
In this way, the limits of the $\varphi$ integration are
$\pm \varphi_{12}/2$. The lower limit of the $\theta$ integration is
zero, whereas the upper limit depends on $\varphi$ and varies from
$\tau_0$ (at $\varphi=\pm \varphi_{12}/2$) to $\eta_{12}$ (at 
$\varphi=0$).
If we use, instead of $\theta$, the variable $\xi$, its limits
will be $0\leq \xi \leq 1$, independently of $\varphi$.
The result of an infinitesimal variation of $\xi$ will be
a thin strip based on the side $1_{\xi}2_{\xi}$ (cf. Fig.~7).
Moreover, all such strips from the triangles 012, 023 and 031
(cf. Fig.~6) can be ``glued'' together. So, the whole result of
a variation in $\xi$ is a thin closed spherical triangle.
Alternatively, one can perform the azimuthal $\varphi$-integration
first, and then evaluate the remaining $\xi$ integral.
As we shall see, both ways work in the three-point case.   

Using formulae of spherical trigonometry (cf. Fig.~7), we get
the following relations:
\begin{eqnarray*}
\sin\left({\textstyle{1\over2}}\tau(\xi)\right) 
   &=& \sin\left({\textstyle{1\over2}}\varphi_{12}\right)\;
       \sin(\tau_0\xi) ,
\\
\sin\eta(\xi) 
   &=& \sin\kappa(\xi)\; \sin(\tau_0\xi) ,
\\
\cos\kappa(\xi)
   &=& \sin\left({\textstyle{1\over2}}\varphi_{12}\right)\;
       \cos\eta(\xi) ,
\\
\tan\kappa(\xi)
&=& \left( \cos(\tau_0\xi) 
           \tan\left({\textstyle{1\over2}}\varphi_{12}\right) 
    \right)^{-1} ,
\\
\tan\eta(\xi)  
   &=& \cos\left({\textstyle{1\over2}}\varphi_{12}\right)\;
       \tan(\tau_0\xi) .
\end{eqnarray*}

The situation for arbitrary $\varphi$ is drawn in Fig.~8.
This is a rectangular spherical triangle with an upper vertex 0
and one of the outgoing sides being the perpendicular
of the length $\eta(\xi)$, as in Fig.~7. This side corresponds
to $\varphi=0$. The angle at the vertex 0 is equal to $\varphi$
and a running value of $\varphi$ is understood.
The side opposite to 0 is perpendicular to
the one mentioned before: $\eta(\xi)$. This is a part of the
corresponding side directed to the point $1_{\xi}$ in Fig.~7,
i.e. the line of constant $\xi$. Its length is denoted as
$\rho(\xi,\varphi)$. The third side or ``hypotenuse'' has length
$\theta(\xi,\varphi)$. The vertex opposite to the $\eta(\xi)$
side is the running point with the coordinates $\theta, \varphi$.
The angle at this vertex is denoted as $\kappa(\xi,\varphi)$.

We want to express $\theta$ in terms of $\xi$ and $\varphi$.
Using the formulae of spherical trigonometry for the triangle
in Fig.~8, analogous to ones written above, it is straightforward
to show that 
\begin{equation}
\label{tan_theta}
\tan\theta(\xi, \varphi) 
= \frac{\tan\eta(\xi)}{\cos\varphi}
= \frac{\cos\left({\textstyle{1\over2}}\varphi_{12}\right)}
       {\cos\varphi}\;
  \tan(\tau_0\xi) .
\end{equation} 
The area is obtained by integrating
\begin{equation}
\sin\theta \; \mbox{d}\theta \; \mbox{d}\varphi 
= -\mbox{d}(\cos\theta) \; \mbox{d}\varphi
\Rightarrow -\frac{\partial\left(\cos\theta(\xi,\varphi)\right)}
                  {\partial\xi} \;
\mbox{d}\xi \; \mbox{d}\varphi ,
\end{equation}
where, according to eq.~(\ref{tan_theta}), 
\begin{equation}
\cos\theta(\xi,\varphi)
= \left( 1 + 
\frac{\cos^2\left({\textstyle{1\over2}}\varphi_{12}\right)}
     {\cos^2\varphi}\;
        \tan^2(\tau_0\xi)  \right)^{-1/2} .
\end{equation}

If we first perform the $\varphi$-integral, this effectively
corresponds to the area of the infinitesimal strip between two
sides of constant $\xi$,
\begin{equation}
\label{strip}
- 2 \mbox{d} \left\{\arctan\left(\cos(\tau_0\xi)\; 
\tan\left({\textstyle{1\over2}}\varphi_{12}\right) \right) \right\} .
\end{equation}
Then, the remaining $\xi$ integral gives just 
this $\arctan$. Collecting the results for all triangles (cf. Fig.~6),
we get the same as in eq.~(\ref{Omega^{(3)}}). This can be
seen if we observe that the expression (\ref{strip})
is nothing but $2\mbox{d}\kappa(\xi)$ and 
the spherical excess for the triangle $01_{\xi}2_{\xi}$
is $(\varphi_{12}+2\kappa(\xi)-\pi)$. 
A nice thing about eq.~(\ref{strip}) is that in the $\xi$-integral 
we can glue all $\arctan$ functions into a symmetric answer
at the {\em infinitesimal} level.
The same final result can also be obtained when we first integrate
over $\xi$, it being trivial since the integrand is a derivative 
with respect to $\xi$.

\subsection{The four-dimensional case}

If we consider the four-dimensional three-point function, the
only (but very essential!) difference is that we should divide the
integrand
by $\cos\theta(\xi,\varphi)$, cf. eq.~(\ref{gen}).
Therefore, the function in the integrand can be written as
\begin{equation}
-\frac{\partial}{\partial\xi} \ln\left(\cos\theta(\xi,\varphi)\right) .
\end{equation}   
As in the previous case, one can integrate first over either $\varphi$
or $\xi$; both ways lead to equivalent results.
Let us integrate first over $\xi$, this gives
\begin{equation}
\ln\left(\frac{\cos\theta(0,\varphi)}{\cos\theta(1,\varphi)}\right)
= {\textstyle{1\over2}} \ln\left( 1 + \frac{\tan^2\eta_{12}}{\cos^2\varphi}\right) .
\end{equation}
In the equal-mass case, the remaining $\varphi$-integral becomes
\begin{eqnarray}
\int\limits_{0}^{\varphi_{12}/2}
\mbox{d}\varphi
\ln\left( 1 + \frac{\tan^2\eta_{12}}{\cos^2\varphi}\right) 
\hspace{90mm}
\nonumber \\
= {\textstyle{1\over2}}\int\limits_{0}^{\varphi_{12}}
\mbox{d}\varphi'
\left\{
\ln\left( 1 + 2\tan^2\eta_{12} \right)
+\ln\left( 1 + \frac{\cos\varphi'}{1 + 2\tan^2\eta_{12}} \right)
-\ln\left( 1 + \cos\varphi' \right)
\right\} ,
\end{eqnarray}
where $\varphi'=2\varphi$. Using a result from \cite{Lewin} 
(p.~308, eq.~(38)),
this integral can be expressed in terms of the Clausen function 
(see in Appendix~A) as
\begin{equation}
\label{Clausen1}
{\textstyle{1\over2}}\tau_{12}\; 
\ln\left(
\frac{\sin\left({\textstyle{1\over2}}(\varphi_{12}+\tau_{12})\right)}
     {\sin\left({\textstyle{1\over2}}(\varphi_{12}-\tau_{12})\right)}
     \right)
+ {\textstyle{1\over2}}\mbox{Cl}_2\left(\varphi_{12}+\tau_{12}\right)
+ {\textstyle{1\over2}}\mbox{Cl}_2\left(\varphi_{12}-\tau_{12}\right)
- \mbox{Cl}_2\left(\varphi_{12}\right) . 
\end{equation}
The arguments of the Clausen functions have a very transparent geometric
interpretation. We note that the logarithmic term can also be 
presented as
\begin{equation}
{\textstyle{1\over2}} \tau_{12} \;
\ln\left( \frac{1+\sin\eta_{12}}{1-\sin\eta_{12}} \right) .
\end{equation}
The function (\ref{Clausen1}) can also
be expressed in terms of the generalized inverse tangent integral 
(\ref{def_Ti2}) as
\begin{equation}
\mbox{Ti}_2\left(\tan\left({\textstyle{1\over2}}\tau_{12}\right),
      \tan\left({\textstyle{1\over2}}\varphi_{12}\right) \right)
-\mbox{Ti}_2\left(\tan\left({\textstyle{1\over2}}\tau_{12}\right),
      -\tan\left({\textstyle{1\over2}}\varphi_{12}\right) \right) .
\end{equation} 

The generalization to the case of unequal masses is easy.
To understand why, let us consider Fig.~9.
This is again a close-up of the triangle
012 from Fig.~6, but now it is asymmetric, because
$\tau_{01}$ and $\tau_{02}$ are different.
We denote the quantities in one
of the triangles as 
$\varphi_{12}^{+}/2, \; \tau_{12}^{+}/2, \; \kappa_{12}^{+}/2$,
and in another triangle as
$\varphi_{12}^{-}/2, \; \tau_{12}^{-}/2, \; \kappa_{12}^{-}/2$.
The height $\eta_{12}$ is the same for both triangles.
Obviously, 
\begin{equation}
{\textstyle{1\over2}}\left(\varphi_{12}^{+}
     +\varphi_{12}^{-}\right)=\varphi_{12}
\hspace{5mm} \mbox{and} \hspace{5mm} 
{\textstyle{1\over2}}\left(\tau_{12}^{+}+\tau_{12}^{-}\right)=\tau_{12}.
\end{equation}
One can immediately see that the result for the integral over this
asymmetric triangle is nothing but a half of the sum of the
functions (\ref{Clausen1}) labelled with plus and with minus.
Namely,
\begin{eqnarray}
\label{Clausen2}
{\textstyle{1\over2}} \tau_{12} \;
\ln\left( \frac{1+\sin\eta_{12}}{1-\sin\eta_{12}} \right) 
&+& {\textstyle{1\over4}}
        \mbox{Cl}_2\left(\varphi_{12}^{+} + \tau_{12}^{+}\right)
+ {\textstyle{1\over4}}
        \mbox{Cl}_2\left(\varphi_{12}^{+} - \tau_{12}^{+}\right)
- {\textstyle{1\over2}}
        \mbox{Cl}_2\left(\varphi_{12}^{+}\right)
\nonumber \\
&+& {\textstyle{1\over4}}
        \mbox{Cl}_2\left(\varphi_{12}^{-} + \tau_{12}^{-}\right)
+ {\textstyle{1\over4}}
        \mbox{Cl}_2\left(\varphi_{12}^{-} - \tau_{12}^{-}\right)
- {\textstyle{1\over2}}
        \mbox{Cl}_2\left(\varphi_{12}^{-}\right) .
\end{eqnarray}
Again, the geometric interpretation of the arguments of the Clausen 
functions is very transparent, thanks to Fig.~9.

Some useful relations are
\begin{eqnarray*}
\cos\left({\textstyle{1\over2}}\tau_{12}^{+}\right)
= \frac{\cos\tau_{01}}{\cos\eta_{12}} ,
\hspace{18mm} 
\cos\left({\textstyle{1\over2}}\tau_{12}^{-}\right)
= \frac{\cos\tau_{02}}{\cos\eta_{12}} ,
\hspace{20mm}
\\
\sin\left({\textstyle{1\over2}}\tau_{12}^{+}\right)
= \sin\tau_{01} \sin\left({\textstyle{1\over2}}\varphi_{12}^{+}\right) ,
\hspace{8mm}
\sin\left({\textstyle{1\over2}}\tau_{12}^{-}\right)
= \sin\tau_{02} \sin\left({\textstyle{1\over2}}\varphi_{12}^{-}\right) ,
\\
\tan\left({\textstyle{1\over2}}\tau_{12}^{+}\right)
= \sin\eta_{12} \tan\left({\textstyle{1\over2}}\varphi_{12}^{+}\right) ,
\hspace{8mm}
\tan\left({\textstyle{1\over2}}\tau_{12}^{-}\right)
= \sin\eta_{12} \tan\left({\textstyle{1\over2}}\varphi_{12}^{-}\right) .
\end{eqnarray*}
Worth noting is
\begin{equation}
\cos\eta_{12} = \frac{m_0 \sqrt{k_{12}^2}}
                     {m_1 m_2 \sin\tau_{12}} .
\end{equation}

To get the complete result for the three-point integral in four
dimensions, we should sum up the functions (\ref{Clausen2})
for all three triangles 012, 023 and 031, and multiply by the
overall factor (see eq.~(\ref{gen})) which is
\begin{equation}
-\mbox{i} \pi^2 \;
\frac{1}{\sqrt{\Lambda^{(3)}}} .
\end{equation}
Note that in this case we get $\sqrt{\Lambda^{(3)}}$ 
rather than $\sqrt{D^{(3)}}$ in the denominator.

\section{Four-point function}
\setcounter{equation}{0}

\subsection{The general case}

In this case, the four-dimensional simplex has four mass sides
and six momentum sides (see in Fig.~3). 
It has five vertices and five three-dimensional
hyperfaces. Four of these hyperfaces are the {\em reduced} ones,
corresponding to three-point functions, whereas
the fifth one is the momentum hyperface.
In fact, this four-dimensional simplex is completely defined by 
its mass sides
$m_1, m_2, m_3, m_4$ and six ``planar'' angles between them,
$\tau_{12}, \tau_{13}, \tau_{14}, \tau_{23}, \tau_{24}$ and $\tau_{34}$.
According to eq.~(\ref{V^{(N)}}), the content (hyper-volume) of this 
simplex equals
\begin{equation}
\label{V4}
V^{(4)} = {\textstyle{1\over{24}}} \; m_1 m_2 m_3 m_4 \; \sqrt{D^{(4)}} ,
\end{equation}
with $D^{(4)}$ given by (\ref{c-matrix})--(\ref{D(N)}) at $N=4$,
\begin{equation}
\label{D4}
D^{(4)}= \det\|c_{jl}\| , 
\hspace{13mm}
\| c_{jl} \| =
\left(
\begin{array}{cccc}
  1     & c_{12} & c_{13} & c_{14}  \\
 c_{12} & 1      & c_{23} & c_{24}  \\
 c_{13} & c_{23} & 1      & c_{34}  \\
 c_{14} & c_{24} & c_{34} & 1 
\end{array}
\right) \;\; .
\end{equation}

The four-dimensional four-point function can be
exhibited as (cf. eq.~(\ref{uniq4}), (\ref{uniq5})):
\begin{equation}
\label{J4}
J^{(4)}(4;1,1,1,1)= {\textstyle{1\over12}}\; \mbox{i} \pi^2 \;
\frac{\Omega^{(4)}}{V^{(4)}}
= \frac{2 \; \mbox{i} \pi^2}{ m_1 m_2 m_3 m_4 } \; 
\frac{\Omega^{(4)}}{\sqrt{D^{(4)}}} .
\end{equation} 
So, the main problem is how to calculate $\Omega^{(4)}$.

In four dimensions, $\Omega^{(4)}$ is the value of the 
four-dimensional generalization of the solid
angle at the vertex of the simplex where all four mass sides meet.
In the spherical case, it can be defined as the volume of a part 
of the unit hypersphere
which is cut out from it by the four three-dimensional 
reduced hyperfaces, each hyperface involving three mass sides
of the simplex.
This hyper-section is a three-dimensional spherical 
tetrahedron\footnote{In the hyperbolic case, this is a hyperbolic
tetrahedron whose volume can be obtained by analytic 
continuation (see below).}
whose six sides (edges) are equal to the angles $\tau_{jl}$.

It is illustrated in Fig.~10 where the $i$-th vertex ($i=1,2,3,4$)
corresponds to the intersection of the $i$-th mass side
of the basic simplex and the unit hypersphere\footnote{It should be
stressed that Fig.~10 is just an illustration (rather than precise
picture) since a realistic non-Euclidean tetrahedron should be
understood as embedded into four-dimensional Euclidean space.}.
Furthermore, the dihedral angles of this {\em three}-dimensional
spherical tetrahedron coincide with those of the 
{\em four}-dimensional basic simplex, $\psi_{jl}$
(cf. eqs.~(\ref{cos_psi}), (\ref{psi_bar})).
The dual matrix $\|\widetilde{c}_{jl}\|$ as well as its determinant
$\widetilde{D}^{(4)}$, cf. eq.~(\ref{D-tilde}),
\begin{equation}
\|\widetilde{c}_{jl}\| = 
\left(
\begin{array}{cccc}
1 & \widetilde{c}_{12} & \widetilde{c}_{13} & \widetilde{c}_{14} \\
\widetilde{c}_{12} & 1 & \widetilde{c}_{23} & \widetilde{c}_{24} \\
\widetilde{c}_{13} & \widetilde{c}_{23} & 1 & \widetilde{c}_{34} \\
\widetilde{c}_{14} & \widetilde{c}_{24} & \widetilde{c}_{34} & 1 
\end{array}
\right) \; ,
\hspace{10mm}
\widetilde{c}_{jl} = -\cos\psi_{jl} \; ,
\hspace{10mm}
\widetilde{D}^{(4)} = \det\|\widetilde{c}_{jl}\| \; ,
\end{equation} 
are usually referred to as the Gram matrix and determinant of the
spherical or hyperbolic tetrahedron.
For example, the dihedral angle at the edge 12 of the spherical
tetrahedron in Fig.~10 is $\psi_{34}$, etc.\footnote{The digits
labelling the edge and the dihedral angle at this edge
should cover the whole set 1234.}
We hote that the dihedral angle is well defined in spherical and
hyperbolic spaces, i.e. it remains unchanged along the given edge.

Unfortunately, there are no {\em simple}
relations like (\ref{solid_angle}) which might make it possible to
express the volume of a spherical (or hyperbolic) tetrahedron 
in terms of its sides or
dihedral angles. In fact, calculation of this volume in an elliptic
or hyperbolic space is a well-known problem of non-Euclidean
geometry (see e.g. in \cite{Lobachevsky,Schlaefli,Coolidge}).
A standard way to solve this problem, say in spherical space,
is to split an arbitrary tetrahedron into a set of birectangular ones.
For example, the tetrahedron 1234 (shown in Fig.~10) is called
{\em birectangular} (or {\em double-rectangular}) if
(i) the edge 12 is perpendicular to the face 234 and 
(ii) the opposite edge, 34, is perpendicular to the face 123.
It is easy to check that in this case three dihedral angles,
$\psi_{13}, \;\; \psi_{14}$ and $\psi_{24}$, are right angles.
The other three are usually denoted \cite{Coxeter,Vinberg} as
\begin{equation}
\label{abg}
\psi_{12} = \alpha, \quad
\psi_{23} = \beta, \quad
\psi_{34} = \gamma .
\end{equation}  
The volume of a birectangular tetrahedron is known (see below) and
can be expressed in terms of Lobachevsky or Schl\"afli functions
which can be related to dilogarithms (see in \cite{Coxeter,Vinberg}).

The volume of the general tetrahedron must be a symmetrical 
function of the six edges $\tau_{jl}$, or equivalently of the six 
dihedral angles $\psi_{jl}$. When we break it up into a sum
of birectangular tetrahedra, this explicit symmetry may be lost,
although of course it must be hidden in the properties of the sums of
dilogarithmic functions. 

The general way to split an arbitrary tetrahedron into a sum of
birectangular ones is to fix a point 0, say inside the 
tetrahedron\footnote{If this point (or the feet of some of the 
perpendiculars) happen to be outside the tetrahedron, this would
mean that some of the volumes of the resulting birectangular
tetrahedra should be taken with an opposite sign.}.
Then, connecting this point 0 with each of the vertices of the
tetrahedron 1234, we split it into four smaller tetrahedra:
0123, 0124, 0134 and 0234. Consider one of these tetrahedra, say 0123.
Dropping the perpendicular from 0 onto the face 123 and connecting the
foot of this perpendicular $F_{123}$ with the vertices
1, 2 and 3, we split 0123 into three {\em rectangular}
tetrahedra, $012F_{123}$, $013F_{123}$ and $023F_{123}$.
Then, each of these rectangular tetrahedra (say, $012F_{123}$)
can be split into a sum of two {\em birectangular}
tetrahedra by dropping a perpendicular from
$F_{123}$ onto the only side belonging to the original
tetrahedron 1234 (the edge 12).
In this way, we get $4\times3\times2=24$ birectangular tetrahedra.

Since we are free to choose the position of the point 0,
there are (at least) two ways to reduce the number of terms
involved in this volume splitting in the general case:

(i) The point 0 may coincide with one of the vertices 
of the original tetrahedron, say with the vertex 4
(cf. Fig.~11a). In this case, we do not need to split
into four smaller tetrahedra and can start just by dropping 
a perpendicular from the point 4 onto the face 123.
In this way, we reduce the number of birectangular tetrahedra 
involved to six. However, the price we pay for this is the loss
of explicit symmetry.

(ii) A symmetric choice of the point 0 can be related to the 
structure of the basic four-dimensional Euclidean simplex. In
particular, we can take as the point 0 the intersection of the 
height $H_0$ and 
the unit hypersphere. Moreover, since the general
four-point function can be reduced to the equal-mass case
(see eqs.~(\ref{eqm1})--(\ref{eqm2})), 
it is enough to consider the case
when the point 0 is equidistant from the vertices 1, 2, 3 and 4.
In this case, at the first stage of splitting we get
four {\em ``isosceles''} tetrahedra, signifying for each
of them the three sides meeting at the point 0 are equal.
Continuing the splitting, we see that due to ``isoscelesness''
the pairs of the tetrahedra produced at the last stage have
equal values. Effectively, we get the sum of $4\times3=12$ volumes
of birectangular tetrahedra.
Although this is twice as much as in the case (i), the advantage
is that we are able to preserve the explicit symmetry at all stages.

We note that the possibility to rescale the Feynman 
parameters\footnote{Cf. eqs.~(\ref{f1})--(\ref{f2}): for $n=N$, the
denominator
disappears. In general, the $f_i$ can take arbitrary positive values.} 
in the case $n=N$ is closely connected with the freedom to choose  
the point 0 as we like. As we have seen, this rescaling preserves 
the angles $\tau_{jl}$ between the mass vectors, whereas the
direction of the height $H_0$ varies. In particular, we can
make it coincide with with one of the mass vectors,
which corresponds to the splitting (i) considered above. From 
this point of view, we lose the explicit symmetry since
the corresponding rescaling is not symmetric.

Another possibility to reduce the number of birectangular volumes
involved is to try to choose the point 0 in such a way that the
volumes of the four tetrahedra (produced at the first stage
of splitting) are equal (or proportional) to each other.
In this case, the further splitting would give just $3\times2=6$
birectangular terms, i.e. the same number as in the case (i);
but here we may preserve the explicit symmetry. However,
it seems to be difficult (if possible at all) to define
the position of the point 0 in this case, because one needs to solve
a set of transcendental equations.

In some special examples described below we show that an
additional symmetry of the diagram can also be employed.

\subsection{Volume of birectangular tetrahedra in curved space}

As we have seen (cf. eq.~(\ref{abg})), the birectangular tetrahedron 
may be specified by three
dihedral angles $\psi_{12}=\alpha, \psi_{23}=\beta, \psi_{34}=\gamma$,
so that the corresponding Gram matrix $\|\widetilde{c}_{jl}\|$
and its determinant (cf. eq.~(20) of \cite{Vinberg}) are
\begin{equation}
\|\widetilde{c}_{jl}\|=
\left(
\begin{array}{cccc}
1 & -\cos\alpha & 0 & 0 \\
-\cos\alpha & 1 & -\cos\beta & 0 \\
0 & -\cos\beta & 1 & -\cos\gamma \\
0 & 0 & -\cos\gamma & 1 
\end{array}
\right) \; , 
\hspace{7mm}
\widetilde{D}^{(4)}= \sin^2\alpha \sin^2\gamma - \cos^2\beta \; .
\end{equation}
 
The volume of this birectangular tetrahedron (i.e. its
contribution to $\Omega^{(4)}$) can be presented  as
\begin{equation}
\label{V-S}
\Omega^{(4)} = V(\alpha,\beta,\gamma) =
{\textstyle{1\over4}} \;
S(\pi/2-\alpha,\beta,\pi/2-\gamma) .
\end{equation}
The notation $S(\alpha,\beta,\gamma)$ corresponds to the
Schl\"afli function as defined in \cite{Coxeter}, 
whereas the notation $V(\alpha,\beta,\gamma)$ was used in
\cite{Vinberg} and is proportional to the function
originally used by Schl\"afli \cite{Schlaefli}.
Note that the function (\ref{V-S}) is symmetric
with respect to $\alpha$ and $\gamma$.

If we define (cf. eq.~(3.21) of \cite{Coxeter})
\begin{equation}
\label{X}
X\equiv\frac{\cos\alpha \cos\gamma - \sqrt{\widetilde{D}^{(4)}}}
       {\cos\alpha \cos\gamma + \sqrt{\widetilde{D}^{(4)}}} \; ,
\end{equation}
the result for the volume can be expressed in terms of 
dilogarithms \cite{Coxeter},
\begin{eqnarray}
\label{V_Li2}
4 V(\alpha,\beta,\gamma) = 
{\textstyle{1\over2}} \left[ 
\mbox{Li}_2\left(Xe^{2i\alpha}\right) + 
\mbox{Li}_2\left(Xe^{-2i\alpha}\right) \right]
+ {\textstyle{1\over2}} \left[ 
\mbox{Li}_2\left(Xe^{2i\gamma}\right) + 
\mbox{Li}_2\left(Xe^{-2i\gamma}\right) \right]
\hspace{10mm}
\nonumber \\
- {\textstyle{1\over2}} \left[ 
\mbox{Li}_2\left(-Xe^{2i\beta}\right) + 
\mbox{Li}_2\left(-Xe^{-2i\beta}\right) \right]
- \mbox{Li}_2\left(-X\right)
-\left( \frac{\pi}{2} - \alpha \right)^2
-\left( \frac{\pi}{2} - \gamma \right)^2
+ \beta^2 .
\end{eqnarray}

When the condition
$\widetilde{D}^{(4)} \geq 0 \;\; (\sin\alpha\sin\gamma\leq\cos\beta)$
is obeyed, the tetrahedron exists in the usual geometrical sense,
and its volume is real. In this case, one can express 
$V(\alpha,\beta,\gamma)$ in terms of the real function
$\mbox{Li}_2(r,\theta)\equiv
\mbox{Re}\left[\mbox{Li}_2\left(re^{i\theta}\right)\right]$
(see in Appendix~A).
However, one should distinguish between the cases 
$X\geq0$ (or $\cos^2\alpha+\cos^2\beta+\cos^2\gamma\geq1$)
and $X\leq0$ (or $\cos^2\alpha+\cos^2\beta+\cos^2\gamma\leq1$).
The corresponding expressions are 
\begin{eqnarray}
\label{V_Li2_1}
\left. 4 V(\alpha,\beta,\gamma)\right|_{X\geq0} =
\mbox{Li}_2\left(X,2\alpha\right) + 
\mbox{Li}_2\left(X,2\gamma\right) - 
\mbox{Li}_2\left(X,\pi-2\beta\right) 
- \mbox{Li}_2\left(-X\right)  
\hspace{10mm}
\nonumber \\
-\left( \frac{\pi}{2} - \alpha \right)^2
-\left( \frac{\pi}{2} - \gamma \right)^2
+ \beta^2 ,
\hspace{10mm}
\end{eqnarray}
\begin{eqnarray}
\label{V_Li2_2}
\left. 4 V(\alpha,\beta,\gamma)\right|_{X\leq0} =
\mbox{Li}_2\left(-X,\pi-2\alpha\right) + 
\mbox{Li}_2\left(-X,\pi-2\gamma\right) 
- \mbox{Li}_2\left(-X,2\beta\right)
- \mbox{Li}_2\left(-X\right)
\nonumber \\
-\left( \frac{\pi}{2} - \alpha \right)^2
-\left( \frac{\pi}{2} - \gamma \right)^2
+ \beta^2 .
\hspace{10mm}
\end{eqnarray}
In particular, when $X=-1$ we take into account that
$\mbox{Li}_2(1,\theta)=
{\textstyle{1\over4}} (\pi-\theta)^2 - {\textstyle{1\over12}}\pi^2$
and get
\begin{equation}
\left. V(\alpha,\beta,\gamma)\right|_{X=-1} 
= {\textstyle{1\over4}} \pi \left( \alpha+\beta+\gamma-\pi \right) .
\end{equation}
When $\alpha=\beta=\gamma={\textstyle{1\over2}}\pi$, this gives 
${\textstyle{1\over8}}\pi^2$, which agrees with eq.~(\ref{special2})
(at $N=4, \;\; \nu_i=1$).  

When 
$\widetilde{D}^{(4)}<0 \;\; (\sin\alpha\sin\gamma<\cos\beta)$,
the tetrahedron does not exist in the usual geometrical sense
(in the spherical case), but its volume (more precisely, the
analytic continuation of the function corresponding to the volume)
can be obtained from eq.~(\ref{V_Li2}), and it is imaginary.
In this case we define (according to
refs.~\cite{Coxeter,Vinberg})
\begin{equation}
X=e^{-2i\delta}, \hspace{7mm}
\mbox {so that} \hspace{7mm}
\tan\delta =
\frac{\sqrt{-\widetilde{D}^{(4)}}}
     {\cos\alpha\cos\gamma},
\end{equation}
and we get
\begin{eqnarray}
\label{V_Cl2}
V(\alpha,\beta,\gamma) = \frac{1}{8\mbox{i}}
\left[ \mbox{Cl}_2\left(2\alpha+2\delta\right) 
     - \mbox{Cl}_2\left(2\alpha-2\delta\right)
     + \mbox{Cl}_2\left(2\gamma+2\delta\right) 
     - \mbox{Cl}_2\left(2\gamma-2\delta\right)
\right. 
\nonumber \\
\left.
     - \mbox{Cl}_2\left(\pi-2\beta+2\delta\right) 
     + \mbox{Cl}_2\left(\pi-2\beta-2\delta\right)
     + 2 \mbox{Cl}_2\left(\pi-2\delta\right)
\right] \; .
\end{eqnarray} 
Using the connection between the Clausen function and the
Lobachevsky function $L(\theta)$ (cf. eq.~(\ref{Cl2-L}) in
Appendix~A),
eq.~(\ref{V_Cl2}) can be re-written in terms of Lobachevsky
function\footnote{The function used in \cite{Vinberg} which was 
denoted by a Cyrillic ``L'' can be expressed in terms of
the Clausen function as 
${\textstyle{1\over2}}\mbox{Cl}_2\left(2\theta\right)$.}.
The same expression (\ref{V_Cl2}) gives the volume of
a birectangular tetrahedron in the hyperbolic space
(see in \cite{Coxeter,Vinberg}). 

\subsection{Some cases with an additional symmetry}

There are a few cases of particular practical interest because of 
their inherent higher symmetries. In particular, there is the most 
symmetrical case where all $c_{ij}= c$. Thus all dihedral angles of 
the full (regular) tetrahedron are equal, to $\psi$ say. In that case 
the six birectangular tetrahedra corresponding to splitting (i) are 
all of equal size and $\Omega^{(4)}$ reduces to
\begin{equation}
\Omega^{(4)} =
6\;V(\pi/3,\psi/2,\psi)
={\textstyle{3\over2}}\;S(\pi/6,\psi/2,\pi/2-\psi).
\end{equation}
Coxeter gives the identities (eqs.~(4.31)--(4.32) of \cite{Coxeter}) 
\begin{equation}
\label{Coxeter}
S(\pi/6,\psi/2,\pi/2-\psi) = 4S(\pi/6,\pi/3,(\pi-\psi)/2)
 = {\textstyle{2\over3}}\;S((\pi-\psi)/2,\psi,(\pi-\psi)/2),
\end{equation}
which he derived algebraically, and which can be used to convert 
$\Omega^{(4)}$ given by eq.~(\ref{V-S}) into other forms. We will
presently explain the geometrical meaning of these identities.

Another case of interest corresponds to equal mass scattering with equal 
internal masses: this means $c_{12}=c_{23}=c_{34}=c_{14}$ and in general 
are not equal to $c_{13}, c_{24}$, which are associated with energy and 
scattering angle variables. This case has four of the tetrahedral sides 
equal and different from the other two (which are opposite to one 
another). Thus there are three distinct $c$'s. We shall see that it is 
possible here to break up full tetrahedron into four birectangular ones 
by joining the midpoints of the unequal sides to render the problem more 
tractable and make the maximum use of what symmetry exists.

We shall carry out our analysis by relaxing the symmetry relations. 
First we consider the case where all six $c_{ij}$ are equal to $c$ and 
treat this in three different ways. Then we let $c_{13}\equiv c_s$ 
and $c_{24}\equiv c_t$
be unequal. 

{\bf Completely symmetrical box diagram.}
Here all $c_{jl}$ are equal and we are dealing with a completely regular
tetrahedron.
For example, in the equal-mass case this would mean that all $k_{jl}^2$
are equal\footnote{Of course this represents an
unphysical point in the Mandelstam diagram.}, to $k^2$ say, and all
$c=1-k^2/(2m^2)$.
We have
\begin{equation}
\label{tilde_c}
\|c_{jl}\|   = \left(\begin{array}{cccc}
               1 & c & c & c\\
               c & 1 & c & c\\
               c & c & 1 & c\\
               c & c & c & 1
              \end{array} \right),
        \quad
\|\widetilde{c}_{jl}\| =
        \left(\begin{array}{cccc}
               1 & \widetilde{c} & \widetilde{c} & \widetilde{c} \\
               \widetilde{c} & 1 & \widetilde{c} & \widetilde{c} \\
               \widetilde{c} & \widetilde{c} & 1 & \widetilde{c} \\
               \widetilde{c} & \widetilde{c} & \widetilde{c} & 1
              \end{array} \right),
 \quad
\widetilde{c} = -\frac{c}{1+2c} .
\end{equation}      
This means that the dihedral angle $\psi$ between each and every pair of 
hyperplanes\footnote{The generalization of this totally symmetric case 
to an $N$-point function is quite easy and leads to 
$\cos\psi=c/[1+(N-2)c]$.}
equals $\arccos(\frac{c}{1+2c})$.

We can evaluate the hypervolume in at least three ways (the first 
two correspond to the ways (i) and (ii) considered in the general case):

(i) By dropping a perpendicular from one corner onto the opposite
hyperplane
and another perpendicular onto one of the opposite sides, we subdivide 
the full tetrahedron into six equal birectangular tetrahedra
(see Fig.~11a). By inspection, each 
birectangular tetrahedron has dihedral angles 
$\psi, \psi/2$ and $\pi/3$.
Therefore the total hypervolume equals
\begin{equation}
\label{Coxeter_i}
\Omega^{(4)} = 6 \; V(\psi,\;\psi/2,\;\pi/3)
={\textstyle{3\over2}}\; 
S(\pi/6,\;\psi/2,\;\pi/2-\psi) .
\end{equation}

(ii) We can alternatively consider the total volume as four times the
volume contained in the tetrahedron generated by one base triangle 
and the ``centre of mass'' point 0: $P_0 = {\textstyle{1\over4}}(P_1+P_2+P_3+P_4)$
(here, we use the notation $P_i=m_i a_i$, cf. eq.~(\ref{sun2})). 
In this case the matrix associated
with the tetrahedron 0123 is found to be 
\begin{equation}
\label{cp}
\|c_{jl}\|   = \left(\begin{array}{cccc}
               1  & c' & c' & c'\\
               c' & 1  & c  & c\\
               c' & c  & 1  & c\\
               c' & c  & c  & 1
              \end{array} \right),
\quad
c' = \frac{\sqrt{1+3c}}{2}
\end{equation}
leading to
\begin{equation}
\label{tilde_cp}
 \|\widetilde{c}_{jl}\|  
= \left(\begin{array}{cccc}
               1 & \widetilde{c}' & \widetilde{c}' & \widetilde{c}' \\
               \widetilde{c}' & 1 & 1/2 & 1/2 \\
               \widetilde{c}' & 1/2 & 1 & 1/2 \\
               \widetilde{c}' & 1/2 & 1/2 & 1
               \end{array} \right) ,
\quad
\widetilde{c}' = - \sqrt{\frac{1+3c}{2(1+2c)}} . 
\end{equation}
This implies that there are two distinct dihedral angles, 
\begin{equation}
{\textstyle{2\over3}}\pi \hspace{6mm} \mbox{and}
\hspace{6mm}
\arccos\sqrt{\frac{1+3c}{2(1+2c)}}
= {\textstyle{1\over2}} \arccos\left(\frac{c}{1+2c}\right) 
=\frac{\psi}{2}
\end{equation}
in fact. Actually the result is 
hardly surprising since it can be deduced by simple inspection of the 
geometrical figure.
Anyhow if we now divide 0123 into six identical birectangular 
tetrahedra by dropping a perpendicular from 0 onto the centre of 
the triangle 123, these 
smaller tetrahedra possess dihedral angles $\psi/2,\pi/3,\pi/3$. 
Consequently we have an alternative geometrical expression for the full 
volume,
\begin{equation}
\label{Coxeter_ii}
\Omega^{(4)} = 24 \;
V(\pi/3,\;\pi/3,\;\psi/2) 
= 6\; S(\pi/6,\;\pi/3,\;(\pi-\psi)/2).
\end{equation}
Comparing this with the previous answer we have a geometrical 
explanation of one of the Coxeter identities (\ref{Coxeter}). 
Furthermore, because (see \cite{Coxeter}, p.~18) 
there exists the differential relation
\begin{equation}
\mbox{d}S(\pi/6,\pi/3,\gamma)
=-2\arccos\left(\frac{\cos\gamma}{\sqrt{4\cos^2\gamma-1}}\right) 
\mbox{d}\gamma,
\end{equation}
we get the following integral representation of the four-dimensional
solid angle (\ref{Coxeter_ii}):
\begin{equation}
\Omega^{(4)} = \pi^2 - 12\int\limits_0^{(\pi-\psi)/2}
\mbox{d}\gamma \;
\arccos\left(\frac{\cos\gamma}{\sqrt{4\cos^2\gamma-1}}\right) .
\end{equation}

(iii) We take the two midpoints $S$ and $T$ of the 
opposite sides: $P_S=(P_1+P_3)/2, \;\; P_T=(P_2+P_4)/2$
(see in Fig.~11b). 
The full volume is then 
just four times that of the tetrahedron $12ST$, which is already
birectangular as can be seen by inspection. In this case 
the matrix $\|c_{jl}\|$ associated with the tetrahedron $12ST$ 
has the following non-diagonal elements:
\begin{equation}
c_{12}=c, \;\; c_{13}=c_{24}=\sqrt{\frac{1+c}{2}}, \;\;
c_{14}=c_{23}=c\;\sqrt{\frac{2}{1+c}}, \;\;
c_{34}=\frac{2c}{1+c} .
\end{equation}
The corresponding Gram matrix of the tetrahedron $12ST$ is
\begin{equation}
 \|\widetilde{c}_{jl}\|
= \left(\begin{array}{cccc}
               1 & 0 & \widetilde{c}' & 0 \\ 
               0 & 1 & 0 & \widetilde{c}' \\
               \widetilde{c}' & 0 & 1 & \widetilde{c} \\
               0 & \widetilde{c}' & \widetilde{c} & 1
               \end{array} \right) ,
\end{equation}
with $\widetilde{c}$ and  $\widetilde{c}'$ defined in 
eqs.~(\ref{tilde_c}) and (\ref{tilde_cp}), respectively. From 
this we extract the three dihedral angles,
\begin{equation}
\alpha =\gamma={\textstyle{1\over2}}\psi \quad \mbox{and} \quad
\beta = \psi ,
\end{equation}
where $\psi$ is nothing but the dihedral angle at the side 12, i.e. 
one of the dihedral angles of
the full tetrahedron. This accords completely with the shape of $12ST$. 
Hence the total volume of the complete tetrahedron $1234$ is given by
\begin{equation}
\Omega^{(4)} = 4V(\psi/2,\psi,\psi/2) =
S\left( (\pi-\psi)/2,\psi,(\pi-\psi)/2\right) .
\end{equation}
Comparing with the earlier expression, this provides a geometric 
meaning of the second Coxeter identity (\ref{Coxeter}). 

{\bf Partially symmetric box diagram.}
In this case we shall take four sides of the tetrahedron to be equal 
and two opposite sides to be unequal to them. In the equal-mass
case, this would correspond to setting 
$k_{12}^2=k_{23}^2=k_{34}^2=k_{14}^2=k^2$, 
$k_{13}^2 = s, \;\;k_{24}^2=t$, and we would get three different $c$'s:
$c=1-k^2/2m^2, \; c_s=1-s/2m^2, \; c_t=1-t/2m^2$
($s$ and $t$ are just the Mandelstam variables).
In this case
much the simplest way of evaluating the volume of the tetrahedron is to
break it up into four equal birectangular tetrahedra by taking 
the midpoints
of the sides 13 and 24, as before: $P_S=(P_1+P_3)/2, P_T=(P_2+P_4)/2.$
There is enough symmetry in the problem that the full tetrahedron 
is four times the birectangular tetrahedron $12ST$. 
The latter is described by the matrix $\|c_{jl}\|$ with the
non-diagonal elements
\begin{eqnarray}
c_{12}=c, \quad c_{13}=\sqrt{\frac{1+c_s}{2}}, 
\quad c_{24}=\sqrt{\frac{1+c_t}{2}}, 
\hspace{20mm}
\nonumber \\
c_{23}=c\;\sqrt{\frac{2}{1+c_s}}, \quad
c_{14}=c\;\sqrt{\frac{2}{1+c_t}}, \quad
c_{34}=\frac{2c}{\sqrt{(1+c_s)(1+c_t)}} .
\end{eqnarray}
The non-diagonal elements of the ``dual'' matrix 
$\|\widetilde{c}_{jl}\|$ are
\begin{eqnarray}
\widetilde{c}_{13}=
-\sqrt{\frac{(1+c_s)(1+c_t)-4c^2}{2(1+c_t-2c^2)}}, \quad
\widetilde{c}_{24}=
-\sqrt{\frac{(1+c_s)(1+c_t)-4c^2}{2(1+c_s-2c^2)}}, 
\nonumber \\
\widetilde{c}_{34}=
-c\;\sqrt{\frac{(1-c_s)(1-c_t)}{(1+c_s-2c^2)(1+c_t-2c^2)}}, \quad
\widetilde{c}_{12}=\widetilde{c}_{14}=\widetilde{c}_{23}=0 .
\end{eqnarray}
We can now read off the three non-trivial dihedral angles,
\begin{eqnarray}
\alpha=\arccos\left(
              \sqrt{\frac{(1+c_s)(1+c_t)-4c^2}{2(1+c_t-2c^2)}} \right),
\quad
\gamma=\arccos\left(
               \sqrt{\frac{(1+c_s)(1+c_t)-4c^2}{2(1+c_s-2c^2)}} \right),
\nonumber \\
\beta=\arccos\left( 
        c\sqrt{\frac{(1-c_s)(1-c_t)}{(1+c_s-2c^2)(1+c_t-2c^2)}} \right).
\hspace{30mm}
\end{eqnarray}
Finally then the required volume of the full tetrahedron 1234 is 
obtained from
\begin{equation}
\Omega^{(4)} =
4\;V(\alpha,\beta,\gamma)=S(\pi/2-\alpha,\beta,\pi/2-\gamma).
\end{equation}

A well-known physical example of partially symmetric box diagram
is related to the photon-photon scattering \cite{photon-photon}.
In this case, $k^2=0$, $c=1$, $\cos\beta=1$, $\beta=0$.
Furthermore, $\widetilde{D}^{(4)}=\sin^2\alpha\sin^2\gamma-1$
and we should use
the formula (\ref{V_Cl2}) for $V(\alpha,0,\gamma)$. 

\subsection{The massless limit}

In the case when all masses $m_i$
vanish, the quantities $c_{jl}$ become infinite (cf. eq.~(\ref{def_c}))
and should be be considered as hyperbolic cosines. 
The determinant $D^{(4)}$, eq.~(\ref{D4}), has the following
limit:
\begin{eqnarray}
\label{massless_D4}
\left( \left. m_1^2 m_2^2 m_3^2 m_4^2 \; D^{(4)} \right) 
\right|_{m_i\to 0}
\hspace{106mm}
\nonumber \\
\Rightarrow {\textstyle{1\over{16}}}
\left[ (k_{12}^2 k_{34}^2)^2 \!+\! (k_{13}^2 k_{24}^2)^2               
       \!+\! (k_{14}^2 k_{23}^2)^2
       \!-\! 2 k_{12}^2 k_{34}^2 k_{13}^2 k_{24}^2
       \!-\! 2 k_{12}^2 k_{34}^2 k_{14}^2 k_{23}^2
       \!-\! 2 k_{13}^2 k_{24}^2 k_{14}^2 k_{23}^2
\right]
\nonumber \\
= {\textstyle{1\over{16}}}\; 
\lambda\left(k_{12}^2 k_{34}^2, k_{13}^2 k_{24}^2, k_{14}^2 k_{23}^2
\right) \hspace{50mm} ,
\end{eqnarray} 
with $\lambda(x,y,z)$ defined by eq.~(\ref{Kallen}).

The elements of the dual matrix $\|\widetilde{c}_{jl}\|$
still can be interpreted as cosines of the dihedral angles,
namely:
\begin{eqnarray}
\label{ideal_da1}
\widetilde{c}_{12}=\widetilde{c}_{34}
= -\cos\psi_{34} = -\cos\psi_{12}
= \frac{k_{13}^2 k_{24}^2 + k_{14}^2 k_{23}^2 - k_{12}^2 k_{34}^2}
       {\sqrt{k_{13}^2 k_{24}^2 k_{14}^2 k_{23}^2}} ,
\\
\label{ideal_da2}
\widetilde{c}_{13}=\widetilde{c}_{24}
= -\cos\psi_{24} = -\cos\psi_{13}
= \frac{k_{14}^2 k_{23}^2 + k_{12}^2 k_{34}^2 - k_{13}^2 k_{24}^2}
       {\sqrt{k_{14}^2 k_{23}^2 k_{12}^2 k_{34}^2}} ,
\\
\label{ideal_da3}
\widetilde{c}_{14}=\widetilde{c}_{23}
= -\cos\psi_{23} = -\cos\psi_{14}
= \frac{k_{12}^2 k_{34}^2 + k_{13}^2 k_{24}^2 - k_{14}^2 k_{23}^2}
       {\sqrt{k_{12}^2 k_{34}^2 k_{13}^2 k_{24}^2}} .
\end{eqnarray}

Therefore, in this situation we get nothing but an {\em ideal}
hyperbolic tetrahedron (see pp.~39--40 of \cite{Vinberg}), i.e.
the tetrahedron whose vertices are all at infinity. The pairs
of opposite dihedral angles of the ideal tetrahedron are equal,
$\psi_{12}=\psi_{34}, \;\; \psi_{13}=\psi_{24}, \;\;
\psi_{14}=\psi_{23}$,
whereas its volume is given by 
\begin{equation}
\label{ideal}
\Omega^{(4)} = \frac{1}{2\mbox{i}} 
\left[ \mbox{Cl}_2\left(2\psi_{12}\right) 
     + \mbox{Cl}_2\left(2\psi_{13}\right) 
     + \mbox{Cl}_2\left(2\psi_{23}\right)
\right] ,
\hspace{13mm}
\psi_{12}+\psi_{13}+\psi_{23}=\pi , 
\end{equation} 
according to eq.~(41) of \cite{Vinberg} (see also in \cite{Milnor}).

First of all we see that eqs.~(\ref{massless_D4})--(\ref{ideal_da3})
depend only on the products $k_{12}^2 k_{34}^2$, $k_{13}^2 k_{24}^2$
and $k_{14}^2 k_{23}^2$. This corresponds to the ``glueing'' of
arguments in the massless case observed in ref.~\cite{UD1-2}.
Moreover, the result (\ref{ideal}) is of the same form as
the massless three-point function in four dimensions 
(\ref{massless_triangle}).
This accords completely with the reduction of the massless 
four-point function to the three-point function which has
been proven in \cite{UD1-2}.

\section{Reduction of integrals with $N>n$ and related problems}
\setcounter{equation}{0}

\subsection{Integrals with $N>n$}

Consider eq.~(\ref{split2_JN}) in the case when $n=N-1$,
\begin{equation}
\label{split3_JN}
J^{(N)}(N-1; 1, \ldots, 1) = \frac{1}{\Lambda^{(N)}}\;
\left(\prod m_i^2 \right) \;
\sum_{i=1}^N \frac{1}{m_i^2} \; F_i^{(N)} \;
J_i^{(N)}(N-1; 1, \ldots, 1) .
\end{equation}
The integrals on the r.h.s., $J_i^{(N)}(N-1; 1, \ldots, 1)$,
correspond to splitting the basic $N$-dimensional simplex 
into $N$ rectangular
ones. The $i$-th rectangular simplex is obtained via
substituting the $i$-th mass side of the basic simplex 
by its height $H_0$ whose length is $m_0$. Therefore, the corresponding
elements of the $\|c\|$ matrix (\ref{c-matrix}) should be
replaced by (cf. eqs.~(\ref{mas_i})--(\ref{mom_i}))
\begin{equation}
c_{ji} \rightarrow \cos\tau_{0j} = \frac{m_0}{m_j},
\hspace{10mm}
c_{il} \rightarrow \cos\tau_{0l} = \frac{m_0}{m_l} .
\end{equation}

At this point let us consider what the representation (\ref{Fp4})
gives for $J_i^{(N)}$.
The $\alpha$-integrand becomes
\begin{equation}
\label{Fp4+}
\left( \sum_{l\neq i}^{}\frac{\alpha_l}{m_l}
       + \frac{\alpha_i}{m_0} \right)
\delta\left(
\left. \left(\alpha^T \|c\| \alpha\right)\right|_{\alpha_i=0}
+ 2 m_0 \alpha_i \sum_{l\neq i}^{}\frac{\alpha_l}{m_l}
+ \alpha_i^2 \!-\! 1 \right) .
\end{equation}
One can see that the first factor in (\ref{Fp4+}) is proportional to
the derivative of the argument of the $\delta$ function 
with respect to $\alpha_i$. Therefore, eq.~(\ref{Fp4+})
can be expressed as
\begin{equation}
\label{Fp4++}
\frac{1}{2 m_0} \; \frac{\partial}{\partial \alpha_i}
\theta\left(
\left. \left(\alpha^T \|c\| \alpha\right)\right|_{\alpha_i=0}
+ 2 m_0 \alpha_i \sum_{l\neq i}^{}\frac{\alpha_l}{m_l}
+ \alpha_i^2 \!-\! 1 \right) .
\end{equation}
Integrating over $\alpha_i$, we get
\begin{equation}
\label{Fp4+++}
J_i^{(N)}(N-1;1,\ldots,1) =
\frac{\mbox{i}^{1-2N}\;\pi^{(N-1)/2}}
     {m_0^2 \prod\limits_{l\neq i}^{} m_l}
\Gamma\left(\frac{N+1}{2}\right)
\int\limits_0^{\infty} \ldots \int\limits_0^{\infty}
\prod_{l\neq i}^{} \mbox{d}\alpha_l \;
\theta\left( 1- \left.
\left(\alpha^T \|c\| \alpha\right)\right|_{\alpha_i=0} \right) .
\end{equation}
Inserting
\begin{equation}
\frac{1}{N-1} \sum\limits_{l\neq i}^{}
\frac{\partial\alpha_l}{\partial\alpha_l}
= 1
\end{equation}
and integrating by parts, the integrand of (\ref{Fp4+++})
can be transformed as

\begin{eqnarray}
\label{chain}
\theta\left( 1- \left.
\left(\alpha^T \|c\| \alpha\right)\right|_{\alpha_i=0} \right) 
& \Rightarrow &
- \frac{1}{N-1} \left( \sum\limits_{l\neq i}^{}   
\alpha_l \frac{\partial}{\partial\alpha_l} \right)
\theta\left( 1- \left.
\left(\alpha^T \|c\| \alpha\right)\right|_{\alpha_i=0} \right)
\nonumber \\
& \Rightarrow &
\frac{1}{N-1}
\delta\left( 1- \left.
\left(\alpha^T \|c\| \alpha\right)\right|_{\alpha_i=0} \right)
 \sum\limits_{l\neq i}^{}
\alpha_l \frac{\partial}{\partial\alpha_l}
\left.\left(\alpha^T \|c\| \alpha\right)\right|_{\alpha_i=0}
\nonumber \\
& \Rightarrow &
\frac{2}{N-1}
\delta\left( \left.
\left(\alpha^T \|c\| \alpha\right)\right|_{\alpha_i=0} -1 \right) .
\end{eqnarray} 

Using eq.~(\ref{Fp4}), we arrive at
\begin{equation}
J_i^{(N)}(N-1;1,\ldots,1)
= - \frac{1}{2 m_0^2} \;
\left. J^{(N-1)}(N-1;1,\ldots,1)
\right|_{\mbox{\scriptsize without}\;i}.
\end{equation}
Therefore, we get
\begin{equation}
\label{reduction1}
J^{(N)}(N-1; 1, \ldots, 1)
= -\frac{1}{2\; D^{(N)}}
\sum\limits_{i=1}^N \frac{1}{m_i^2} \; F_i^{(N)} \;
\left.
J^{(N-1)}(N-1;1,\ldots, 1)\right|_{\mbox{\scriptsize without}\; i} .
\end{equation}
This corresponds to eq.~(47) of \cite{Nickel},
when the number of external legs $N$ is equal to the space-time dimension
plus one. For $N=3,4,5$, eq.~(\ref{reduction1}) reproduces
eqs.~(10), (8), (13) from \cite{Nickel}, respectively\footnote{For
$N=3$, the explicit result was presented in \cite{KT}.}.

Similar reduction formulae have also been considered in
refs.~\cite{Petersson}--\cite{BDK}.
In ref.~\cite{Brown} it was shown how the linear dependence 
of external momenta in the case $N\geq6$ and for $n=4$ may
be employed to reduce the $N$-point function to  a set
of $(N-1)$-point ones. 
A method for expressing the five-point function in four
dimensions in terms of a sum of four-point functions was
first shown in \cite{Halpern}, whereas the reduction of the
three-point function in two dimensions was studied in \cite{KT}
(see also in \cite{KW}). 
These results were generalized in refs.~\cite{Petersson,Melrose}.
For example, in ref.~\cite{Petersson} (see also in \cite{KT}) 
the representation for
the coefficients corresponding to $F_l^{(N)}$ (eq.~(7) of
\cite{Petersson}) is given in terms of the propagators whose
momenta are solutions to a certain set of equations corresponding
to the other propagators.
Moreover, it is possible to make a direct as opposed to step-by-step
reduction to $n$-point functions for any $N>n$. 
More recently, some other approaches to this problem
were suggested in \cite{vNV,BDK}.

It is also instructive to consider another derivation of
eq.~(\ref{reduction1}). The representation (\ref{Fp4}) yields
\begin{equation}
\label{Fp4_again}
J^{(N)}(N\!-\!1;1,\ldots,1)
=\frac{2\mbox{i}^{1-2N}\pi^{(N-1)/2}}{\prod m_l}\;
\Gamma\left(\frac{N\!+\!1}{2}\right)
\int\limits_0^{\infty} \ldots \int\limits_0^{\infty}
\prod\mbox{d}\alpha_l
\left( \sum\limits_{i=1}^{N}\frac{\alpha_i}{m_i} \right) 
\delta\left(\alpha^T\|c\|\alpha\!-\!1\right) .
\end{equation}
Using eq.~(\ref{linear}) we get
\begin{equation}
\label{linear3}
\sum\limits_{i=1}^{N} \frac{\alpha_i}{m_i}
= \frac{1}{D^{(N)}} 
\sum\limits_{i=1}^{N} \alpha_i 
\sum\limits_{l=1}^{N} \frac{c_{il}}{m_l} F_l^{(N)}
= \frac{1}{D^{(N)}} 
\sum\limits_{i=1}^{N} \frac{F_i^{(N)}}{m_i}
\sum\limits_{l=1}^{N} c_{il} \alpha_l .
\end{equation}

Repeating the steps (\ref{Fp4+})--(\ref{chain}) for the resulting
$\alpha$-integrals, we arrive at the following result:
\begin{eqnarray}
\label{without_i}
\frac{2\mbox{i}^{1-2N}\pi^{(N-1)/2}}{\prod m_l}\;
\Gamma\left(\frac{N\!+\!1}{2}\right)
\int\limits_0^{\infty} \ldots \int\limits_0^{\infty}
\prod\mbox{d}\alpha_l
\left( \sum\limits_{l=1}^{N} c_{il} \alpha_l \right)
\delta\left(\alpha^T\|c\|\alpha\!-\!1\right)
\nonumber \\
= -\frac{1}{2m_i}\; \
\left. J^{(N-1)}(N-1;1,\ldots,1)\right|_{\mbox{\scriptsize without}\;i}. 
\end{eqnarray}
Substituting (\ref{linear3}) and (\ref{without_i}) into 
(\ref{Fp4_again}), we obtain once again eq.~(\ref{reduction1}).

Similar approach can also be used to obtain formulae connecting
the integrals with different values of the space-time
dimension \cite{BDK,BM,Tarasov}.

\subsection{Integrals with higher powers of propagators}

Equations like (\ref{without_i}) can also be used for calculating
integrals with higher powers of propagators.
Such integrals can be associated with derivatives with respect
to $m_i^2$, provided that $m_i$ and $k_{jl}^2$ are considered
as the set of independent variables. Specifically, consider the 
sum\footnote{The notation for the set of arguments 
$\{\nu_j\}=\{1+\delta_{lj}\}$
means that all powers of denominators are equal to one,
except for the $\nu_l$ which is equal to two.}
\begin{eqnarray}
\label{sum2}
c_{1i} m_1  J^{(N)}(N+1;2,1,1,\ldots,1)
+ c_{2i} m_2  J^{(N)}(N+1;1,2,1,\ldots,1)
\hspace{30mm}
\nonumber \\
+ \; \ldots \; + c_{iN} m_N  J^{(N)}(N+1;1,\ldots,1,2)
\equiv \sum\limits_{l=1}^N c_{il} m_l J^{(N)}(N+1; \{1+\delta_{lj}\})
\end{eqnarray}
(as usually, it is implied that $c_{ji}=c_{ij}$ and $c_{ii}=1$).
Employing the representation (\ref{Fp4}) for all integrals in
(\ref{sum2}), we obtain the same integral as in eq.~(\ref{without_i}),
multiplied by $(-\pi)$. Therefore, we get a system of equations 
which can be displayed in the matrix form,
\begin{equation}
\label{matrix}
\| c_{jl} \| \;
\left( \begin{array}{c} m_1 \; J^{(N)}(N+1;2,1,1,\ldots,1) \\
                        m_2 \; J^{(N)}(N+1;1,2,1,\ldots,1) \\
                        \ldots \\
                        m_N \; J^{(N)}(N+1;1,1,\ldots,1,2)
       \end{array} \right)
= \frac{\pi}{2}
\left(
\begin{array}{c}
m_1^{-1} \left.
J^{(N-1)}(N-1;1,\ldots, 1)\right|_{\mbox{\scriptsize without}\; 1} \\
m_2^{-1} \left.
J^{(N-1)}(N-1;1,\ldots, 1)\right|_{\mbox{\scriptsize without}\; 2} \\
\ldots \\   
m_N^{-1} \left.
J^{(N-1)}(N-1;1,\ldots, 1)\right|_{\mbox{\scriptsize without}\; N}
\end{array} \right) .
\end{equation} 

The matrix inverse to $\|c_{jl}\|$ is given by eq.~(\ref{inverse}).
Applying $\|c_{jl}\|^{-1}$ to both sides, we get\footnote{Similar 
results can also be obtained using the  
integration-by-parts approach \cite{ibp,jpa}.}
\begin{eqnarray}
\label{matrix3}
\left( \begin{array}{c}
\left( m_1/\sqrt{D_{11}^{(N-1)}}\right) \;
  J^{(N)}(N+1;2,1,1,\ldots,1) \\
\left( m_2/\sqrt{D_{22}^{(N-1)}}\right) \;
  J^{(N)}(N+1;1,2,1,\ldots,1) \\ 
\ldots \\
\left( m_N/\sqrt{D_{NN}^{(N-1)}}\right) \;
  J^{(N)}(N+1;1,1,\ldots,1,2)
     \end{array} \right)
\hspace{60mm}   
\nonumber \\   
= \frac{\pi}{2 \; D^{(N)}} \; \|\widetilde{c}_{jl}\| \;
\left( \begin{array}{c}
m_1^{-1} \sqrt{D_{11}^{(N-1)}} \;
\left.      
J^{(N-1)}(N-1;1,\ldots, 1)\right|_{\mbox{\scriptsize without}\; 1} \\
m_2^{-1} \sqrt{D_{22}^{(N-1)}} \;
\left.
J^{(N-1)}(N-1;1,\ldots, 1)\right|_{\mbox{\scriptsize without}\; 2} \\
\ldots \\
m_N^{-1} \sqrt{D_{NN}^{(N-1)}} \;
\left.
J^{(N-1)}(N-1;1,\ldots, 1)\right|_{\mbox{\scriptsize without}\; N}
       \end{array} \right) .
\end{eqnarray}

Using the expression (\ref{cos_psi}) for $\widetilde{c}_{jN}$,
we arrive at
\begin{equation}
\label{JNplus}
J^{(N)}(N+1; \{ 1+\delta_{il} \} )
= \frac{\pi}{2\; m_l\; D^{(N)}}
\sum\limits_{j=1}^N \frac{(-1)^{j+l}}{m_j} \;
D_{jl}^{(N-1)} \;
\left.
J^{(N-1)}(N-1;1,\ldots, 1)\right|_{\mbox{\scriptsize without}\; j} .
\end{equation}

In particular, in the three-point four-dimensional case we end up with
\begin{equation}
J^{(3)}(4;1,1,2) = \frac{\mbox{i}\pi^2 \; \sin{\tau_{12}}}
                        {2 m_1 m_2 m_3^2 \; D^{(3)}} \;
\left[ \tau_{23} \; \cos{\psi_{13}}
                 + \tau_{13} \; \cos{\psi_{23}}
                 + \tau_{12} \right] .
\end{equation}
The results for $J^{(3)}(4;1,2,1)$ and $J^{(3)}(4;2,1,1)$ 
can be obtained by permutation of indices. 
The sum of these three integrals is in agreement with the
result obtained in \cite{KW}.

It is easy to check the consistency of the results (\ref{JNplus}) 
and (\ref{reduction1}). 
Using the standard Feynman parametric representation (or the
representation (\ref{Fp4}) involving the $\delta$ function of the
quadratic form) it is straightforward to show that\footnote{
The generalization of this formula to the case of arbitrary $n$
and the powers of the propagators $\nu_i$ can be found e.g. in
\cite{PLB'91}, eq.~(6).}
\begin{equation}
\label{sum}
\sum\limits_{j=1}^N J^{(N)}(N+1; \{ 1+\delta_{ij} \} )
= -\pi \; J^{(N)}(N-1; 1, \ldots, 1) .
\end{equation}
Considering the sum of the integrals (\ref{JNplus}) and employing the
relation (\ref{sum}), we arrive at eq.~(\ref{reduction1}).

\section{Conclusion}
\setcounter{equation}{0}

Let us briefly summarize the main results of the present paper.
We have proved that there is a direct link between Feynman
parametric representation of a one-loop $N$-point function
and the basic simplex in $N$-dimensional Euclidean space.
To show this, we have changed the surface of integration
in the Feynman parametric representation, from linear one
($\sum\alpha_i=1$) to quadratic one ($\alpha^T \|c\|\alpha=1$).
The integral representation involving $\delta$ function of
this quadratic form is given by eq.~(\ref{Fp4}).
 
In the case $N=n$, the result for the Feynman integral turns out to
be proportional to the ratio of an $N$-dimensional solid angle
at the meeting point of the mass sides to the content of 
the $N$-dimensional basic simplex\footnote{Moreover, 
in the case $N=n$ the result for the diagram with unequal
masses can be expressed in terms of the function corresponding
to the equal-mass case.}.  
For example, to reproduce result for the three-dimensional 
three-point function \cite{Nickel} we just require the expression
for the area of a spherical triangle which can easlily
be calculated as the sum of the angles minus $\pi$ (in general,
an analytic continuation of the area function should be taken).

For the four-dimensional four-point 
function, the representation (\ref{Fp4})  provides a very 
interesting connection with the volume 
of the non-Euclidean (spherical or hyperbolic) tetrahedron.
By  splitting the non-Euclidean tetrahedron into birectangular 
ones, the latter can be expressed in terms of dilogarithms
or related functions\footnote{A compact integral representation
for this volume can be found in ref.~\cite{Hsiang}.}.
Different ways of splitting correspond to different
representations of the result for the four-point function. 
This ``freedom'' can be used to construct more compact
representations (cf. refs.~\cite{Wu,'tHV-79,DNS}).

In other cases, such as for example the four-dimensional three-point
function, we get an extra factor in the integrand corresponding
to a power of the cosine of the angle between the height of 
the basic simplex and the vector of integration. 
In general, the height of the basic simplex, $H_0$, plays an 
essential role in calculation of the integrals with $N\neq n$.
It is used to split the basic Euclidean simplex into $N$ rectangular
simplices. When $N=n+1$, each integral corresponding to one
of the resulting rectangular tetrahedra can be reduced
to an $(N-1)$-point function.    
When $N<n$, this splitting also simplifies the calculation of 
separate integrals.

The derivations in this paper have been based on a purely 
geometrical approach to evaluation of $N$-point Feynman diagrams.
In the resulting expressions, all arguments of functions arising
possess a straightforward geometrical meaning in terms of the 
dihedral angles, etc. In particular, this is quite useful for 
choosing the most convenient kinematic variables to describe the 
$N$-point diagrams. We suggest that this approach can serve as 
a powerful tool for understanding the geometrical structure of 
loop integrals with several external legs, as well as the
structure of the corresponding phase-space integrals (see e.g.
in \cite{phase}). More optimistically,
it may shed light on analytical results for higher loops.

\vspace{4mm}

{\bf Acknowledgements.} One of the authors (A.~D.) is grateful
to the Australian Research Council for a grant which was used
to support his visit to Hobart in 1996 (where an essential
part of this work has been done), and to the Department of Physics,
University of Tasmania for hospitality during the visit. 
A.~D.'s research was partly supported by the grants
INTAS--93-0744 and RFBR--96-01-00654.

\vspace{4mm}

{\bf Note-added.}
  After our manuscript has been accepted for a publication,
  the paper~\cite{Wagner-JMP} appeared, 
  dealing with non-relativistic Feynman
  integrals. It cited ref.~\cite{OW} where the relativistic one-loop 
  Feynman
  amplitudes (in Euclidean quantum field theory) were studied and
  expressed in terms of definite integrals over simplices.
  We also note that some representations for the Schl\"afli function
  (similar to those discussed in sec.~6.2) can be found 
  in ref.~\cite{Kellerhals}.

\appendix
\section*{Appendix A: \  Dilogarithm and related functions}
\setcounter{equation}{0}
\renewcommand{\thesection}{A}

Here we present definitions and some representations of the
dilogarithm and related functions. More detailed information
can be found in \cite{Lewin}.

The Euler's dilogarithm is defined by
\begin{equation}
\label{def_Li2}
\mbox{Li}_2(z) 
\equiv - z \int\limits_0^1 \mbox{d}\xi \frac{\ln\xi}{1-\xi z} .
\end{equation}
The Clausen function is related to the imaginary part of the
dilogarithm,
\begin{equation}
\mbox{Cl}_2\left(\theta\right)=\mbox{Im}
\left[\mbox{Li}_2\left(e^{\mbox{\scriptsize{i}}\theta}\right)\right] 
= - \int\limits_0^\theta
\mbox{d}\theta' \ln\left|2\sin\left(
{\textstyle{1\over2}}\theta'\right)\right| ,
\end{equation}
and it can also be represented as
\begin{equation}
\mbox{Cl}_2\left(\theta\right)=-\sin\theta \int\limits_0^1
\frac{\mbox{d}\xi \;\; \ln\xi}{1-2\xi\cos\theta+\xi^2} .
\end{equation}
Note that 
$\mbox{Cl}_2\left(2\pi-\theta\right)=-\mbox{Cl}_2\left(\theta\right)$.

If we consider dilogarithm of a general complex argument,
$\mbox{Li}_2\left(re^{\mbox{\scriptsize{i}}\theta}\right)$, we get
\begin{equation}
\mbox{Re}\left[ 
\mbox{Li}_2\left(r e^{{\mbox{\scriptsize{i}}}\theta}\right) \right]
\equiv \mbox{Li}_2(r,\theta) ,
\end{equation}
\begin{equation}
\mbox{Im}\left[ 
\mbox{Li}_2\left(r e^{{\mbox{\scriptsize{i}}}\theta}\right) \right]
= \omega \ln r + {\textstyle{1\over2}} \left[
\mbox{Cl}_2\left(2\theta\right)
+\mbox{Cl}_2\left(2\omega\right)
+\mbox{Cl}_2\left(2\chi\right) \right],
\end{equation}
with $\tan\omega=r\sin\theta/(1-r\cos\theta)$ and 
$\chi\equiv\pi-\theta-\omega$. We note that
\begin{equation}
{\textstyle{1\over2}} \left[
\mbox{Cl}_2\left(2\theta\right)
+\mbox{Cl}_2\left(2\omega\right)
+\mbox{Cl}_2\left(2\chi\right) \right]
= -\sin\theta\int\limits_0^r \frac{\mbox{d}\eta \;\; \ln\eta}
{1-2\eta\cos\theta+\eta^2} .
\end{equation}
The imaginary part can also be rewritten as
\begin{equation}
\mbox{Im}\left[ 
\mbox{Li}_2\left(r e^{{\mbox{\scriptsize{i}}}\theta}\right) \right]
= \mbox{Ti}_2\left(\tan\omega\right)
- \mbox{Ti}_2\left(\tan\omega,\tan\theta\right)
= -r\sin\theta \int\limits_0^1
\frac{\mbox{d}\xi\;\; \ln\xi}{1-2r\xi\cos\theta+r^2\xi^2} ,
\end{equation}
where $\mbox{Ti}_2(z)$ and $\mbox{Ti}_2(z,a)$ are the ordinary and
the generalized inverse tangent integrals, respectively:
\begin{equation}
\label{def_Ti2}
\mbox{Ti}_2(z)=\int\limits_0^z\frac{\mbox{d}z'}{z'} \arctan z' ,
\hspace{13mm}
\mbox{Ti}_2(z,a)=\int\limits_0^z\frac{\mbox{d}z'}{z'+a} \arctan z' .
\end{equation}
In particular, $\mbox{Ti}_2(z)=\mbox{Ti}_2(z,0)$.

Finally the Lobachevsky function $L(\theta)$ is defined by
\begin{equation}
L(\theta)\equiv -\int\limits_0^{\theta} \mbox{d}\theta' 
\ln\left(\cos\theta'\right), \hspace{13mm}
\left( -{\textstyle{1\over2}}\pi \leq \theta \leq {\textstyle{1\over2}}\pi \right) .
\end{equation}
It is related to Clausen's function by
\begin{equation}
\label{Cl2-L}
\mbox{Cl}_2\left(\theta\right) = - 2L\left(\frac{\pi-\theta}{2}\right)
+ (\pi-\theta)\ln 2 , \hspace{10mm}
L(\theta)= -{\textstyle{1\over2}} 
\mbox{Cl}_2\left(\pi-2\theta\right) + \theta \ln 2 .   
\end{equation}

\appendix
\section*{Appendix B: \  On the rescaling of $\alpha$'s}
\setcounter{equation}{0}
\renewcommand{\thesection}{B}

Consider the transformation 
\begin{equation}
\alpha_i={\cal{G}}(\alpha'_1, \ldots , \alpha'_N) \; \alpha'_i, 
\hspace{10mm}
i= 1, \ldots , N ,
\end{equation}
where the ``scaling'' function ${\cal{G}}$ is the same for 
all $\alpha$'s.
The Jacobian of the transformation is
\begin{equation}
\label{app1}
\mbox{Jacobian} = 
\left( \prod\limits_{j=1}^N \alpha'_j 
\frac{\partial{\cal{G}}}{\partial\alpha'_j} \right) \;
\left| \begin{array}{c}
1+d_1  \;\;\;\; 1 \;\;\;\;  \;\;\ldots\;\; \;\;\;\; 1 \;\;\;\; \\
\;\;\;\; 1 \;\;\;\;  1+d_2  \;\;\ldots\;\; \;\;\;\; 1 \;\;\;\; \\
\ldots\ldots\ldots\ldots\ldots\ldots\ldots \\
\;\;\;\; 1 \;\;\;\; \;\;\;\; 1 \;\;\;\;  \;\;\ldots\;\;  1+d_N
\end{array} \right|,
\hspace{10mm}
       d_i \equiv
{\cal{G}}\left(\alpha'_i\frac{\partial{\cal{G}}}{\partial\alpha'_i}
\right)^{-1} .
\end{equation}
Since the determinant in (\ref{app1}) is equal to
\begin{equation}
\label{app2}
\left( \prod\limits_{i=1}^N d_i \right) \;
\left[ 1 + \frac{1}{d_1} + \frac{1}{d_2} + \ldots + \frac{1}{d_N}
\right] ,
\end{equation} 
the expression for the Jacobian reduces to
\begin{equation}
\label{app3}
\mbox{Jacobian} = 
{\cal{G}}^N\;\left( 1+
\frac{1}{{\cal{G}}}
\sum\limits_{j=1}^N \alpha'_j
\frac{\partial{\cal{G}}}{\partial\alpha'_j}\right)
    =  {\cal{G}}^N\;\left(1 + 
\sum\limits_{j=1}^N \alpha'_j\frac{\partial
(\ln{\cal{G}})}{\partial\alpha'_j}\right).
\end{equation}

If ${\cal{G}}$ is a {\em homogeneous} function of $\alpha$'s of the
order $R$, i.e. 
\begin{equation}
{\cal{G}}(Z\alpha'_1, \ldots, Z\alpha'_N)
= Z^R \; {\cal{G}}(\alpha'_1, \ldots, \alpha'_N) ,
\end{equation}
then
\begin{equation}
\sum\limits_{j=1}^N \alpha'_j   
\frac{\partial{\cal{G}}}{\partial\alpha'_j}
= R \; {\cal{G}}
\hspace{10mm} \mbox{and} \hspace{10mm}
\mbox{Jacobian}= \left[ 1+R \right] \; {\cal{G}}^N .
\end{equation}

In particular, if
${\cal{G}}=Q_r/Q'_{r'}$, where  
$Q_{r}$ is a polyniomial of degree $r$ and $Q'_{r'}$ is 
a polynomial of degree $r'$ in $\alpha'$, 
$\mbox{Jacobian}\rightarrow[1+(r-r')]{\cal{G}}^N$, i.e. $R=r-r'$.
In the text we have encountered the situations when 
$r=r'=1 \;\;(R=0)$ and $r=2,\;\; r'=1 \;\;(R=1)$.


\newpage


\begin{center}
{\large\bf Figure captions}
\end{center}

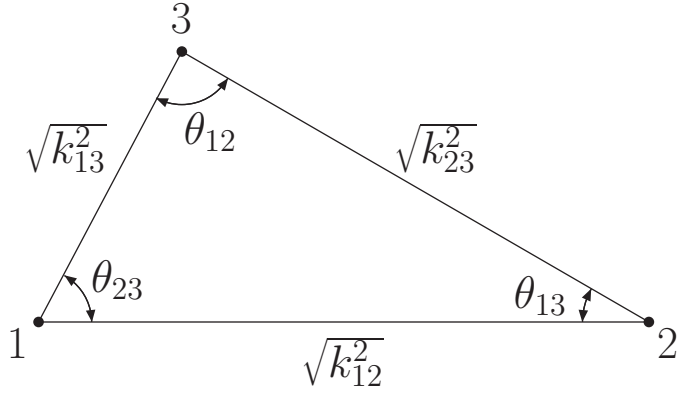
Fig.~1: The triangle associated with the massless 
        three-point function

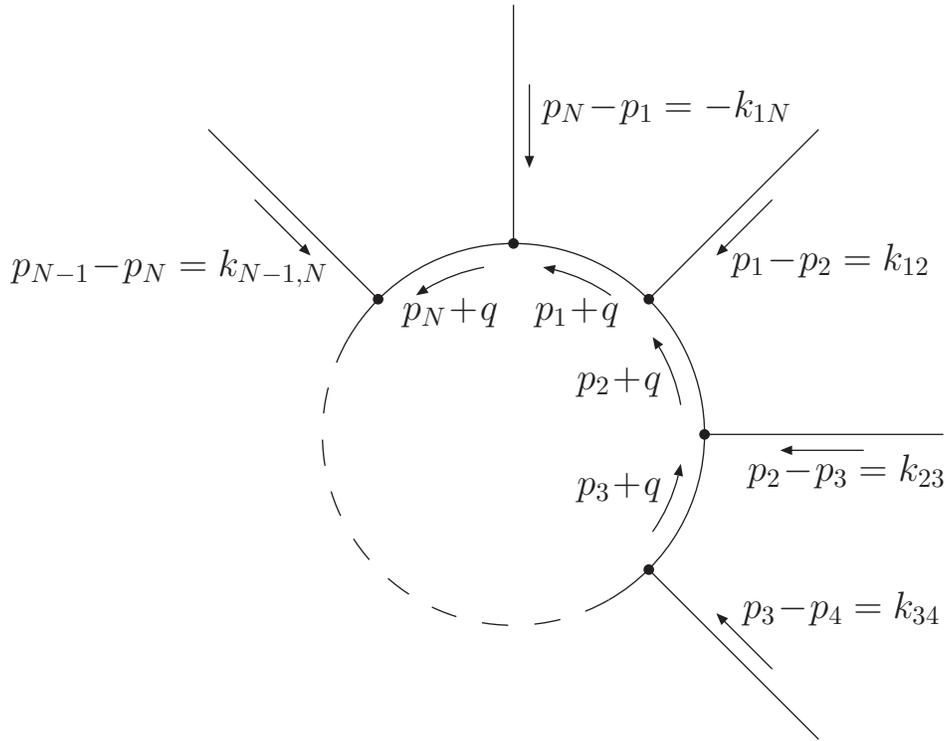
Fig.~2: The one-loop $N$-point diagram

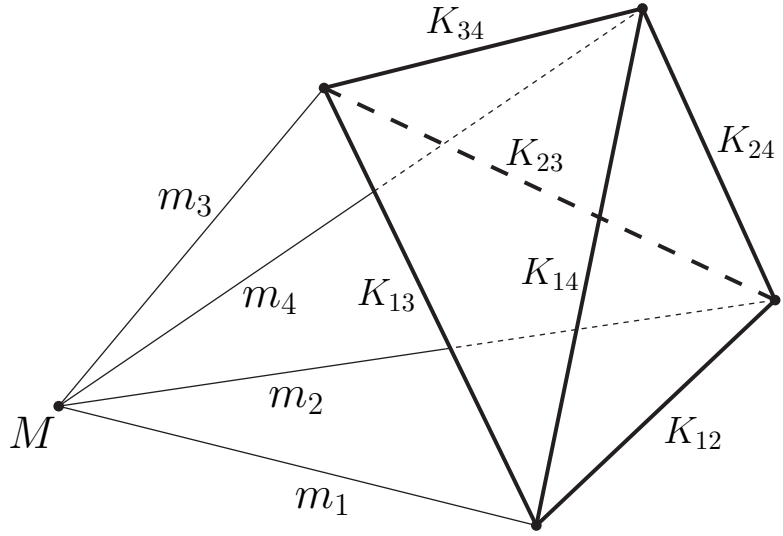
Fig.~3: The basic simplex for $N=4$

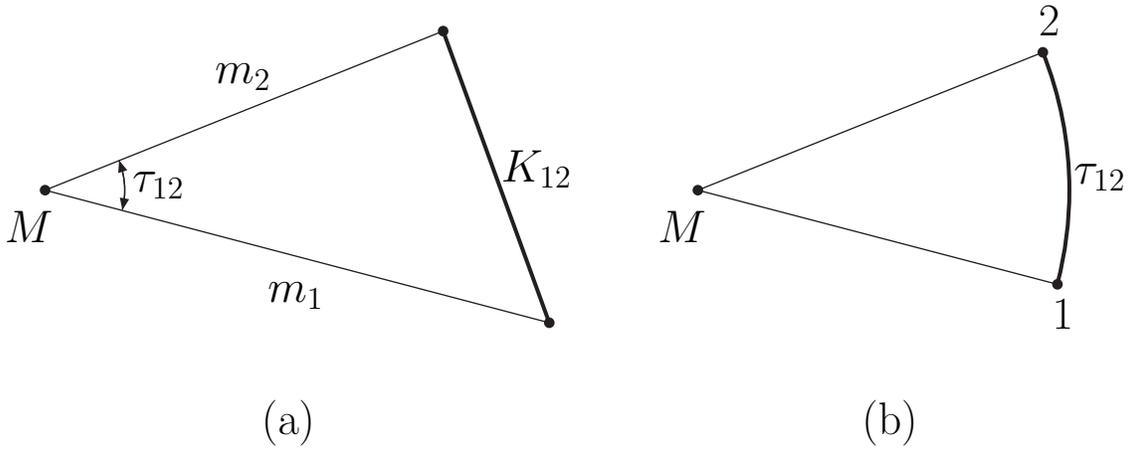
Fig.~4: Two-point case: (a) the basic triangle and 
        (b) the arc $\tau_{12}$ 

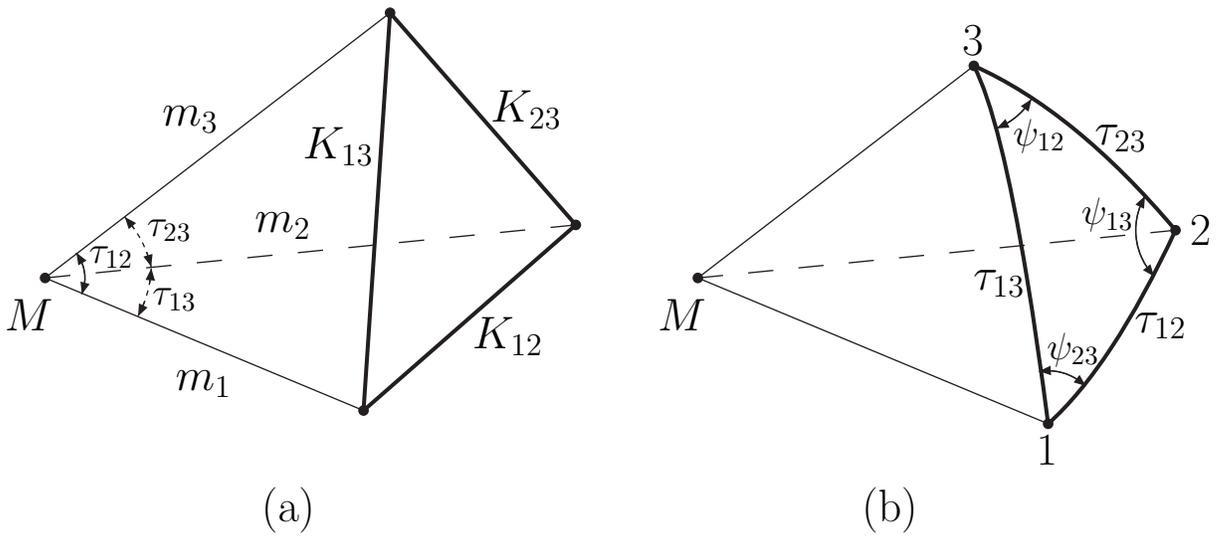
Fig.~5: Three-point case: (a) the basic tetrahedron and
        (b) the solid angle

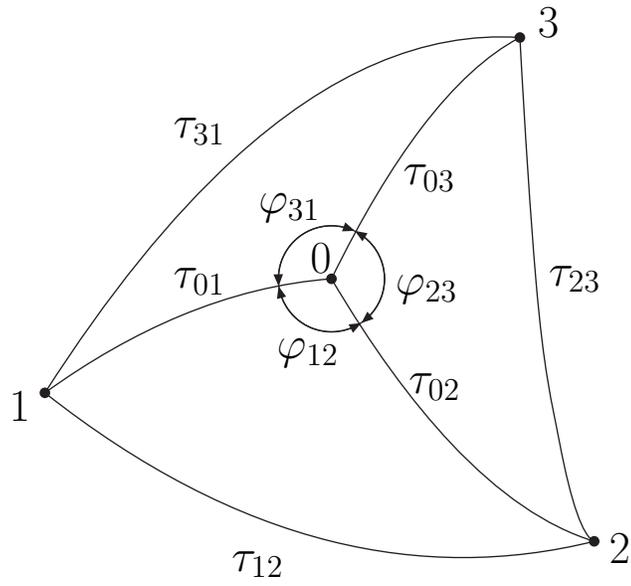
Fig.~6: The spherical triangle 123

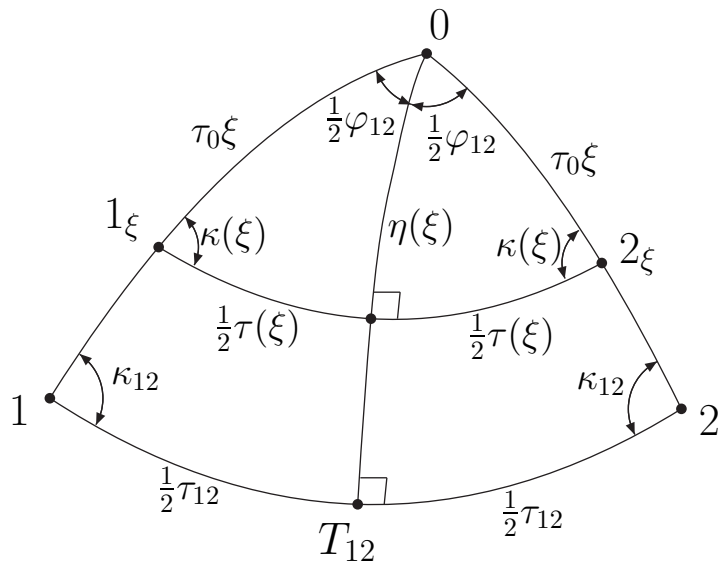
Fig.~7: The isosceles spherical triangle 012

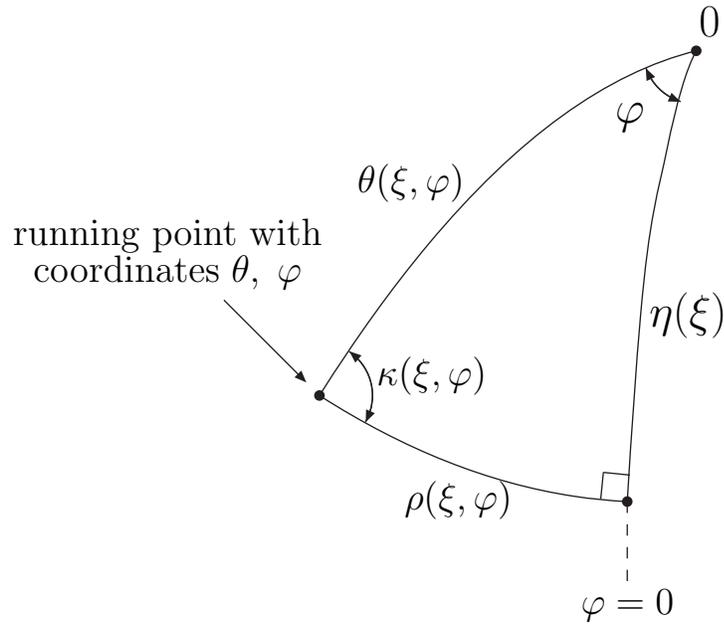
Fig.~8: Integration variables on the sphere

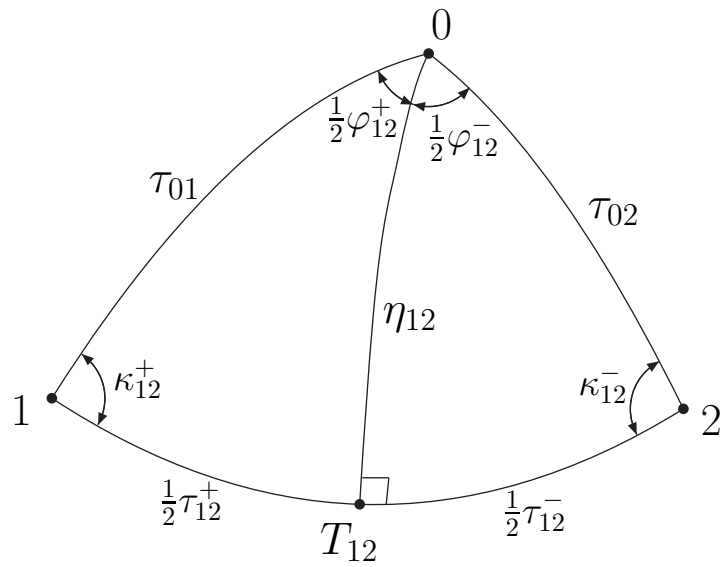
Fig.~9: An asymmetric spherical triangle 012

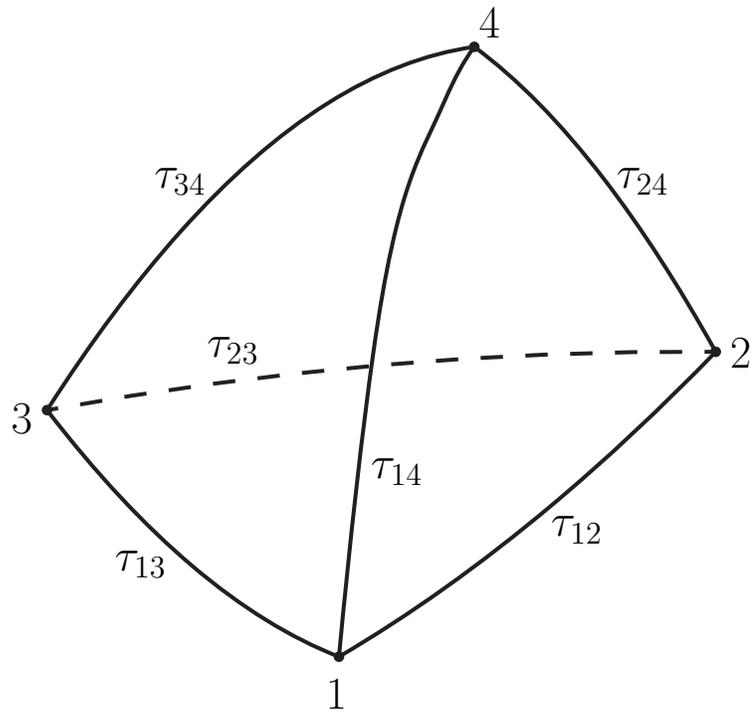
Fig.~10: The spherical tetrahedon 1234

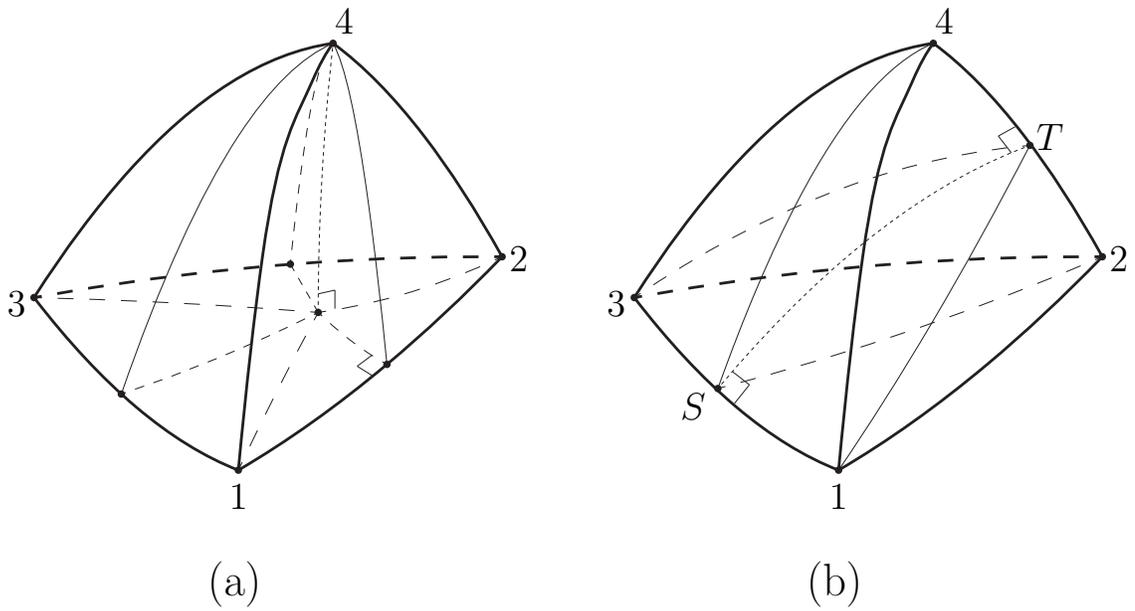
Fig.~11: Different ways of splitting the non-Euclidean tetrahedron

\end{document}